\documentclass[groupedaddress,twocolumn,10pt,pra,aps,showpacs,longbibliography,floatfix,nofootinbib]{revtex4-2}

\usepackage{graphicx}
\usepackage{dcolumn}
\usepackage{bm}
\usepackage{physics}
\usepackage{xcolor}
\usepackage{hyperref}
\usepackage[T1]{fontenc}

\renewcommand{\tr}{\mathrm{tr}}

\newcommand{\m}{\mathbf \Psi}
\newcommand{\w}{\bm \lambda}

\newcommand{\mj}{\m_j}
\newcommand{\wj}{\w_j}

\newcommand{\ukj}{u_k^{(j)}}

\newcommand{\trexpval}[2]{\tr \left( #1^\dagger #2 \right)}
\newcommand{\N}{N}
\newcommand{\NT}{N_T}

\newcommand{\I}{\mathrm{i}}
\newcommand{\diff}[1]{\mathrm{d}#1}

\begin{document}

\preprint{APS/123-QED}

\title{Analytic gradients for low-rank quantum optimal control}%

\author{Leo Goutte}
\email{leo.goutte@epfl.ch}
\author{Vincenzo Savona}%
\email{vincenzo.savona@epfl.ch}
\affiliation{Institute of Physics, Ecole Polytechnique Fédérale de Lausanne (EPFL), CH-1015 Lausanne, Switzerland}
\affiliation{Center for Quantum Science and Engineering, Ecole Polytechnique Fédérale de Lausanne (EPFL), CH-1015 Lausanne, Switzerland}

\date{\today}

\begin{abstract}

We introduce low-rank optimal control (LROC), a method for designing control pulses in open quantum systems whose full density-matrix simulation is prohibitively expensive. The method exploits a feature of quantum computing itself: because protocols are designed to preserve purity, the statistical ensemble described by the density matrix is dominated by a small number of pure states and admits an accurate low-rank factorization. LROC propagates only this factorized form and, by deriving the corresponding adjoint equation, obtains the gradient of any differentiable objective at the same reduced cost as the simulation, leading to a quadratic improvement in time and memory compared to the full master equation. We illustrate the breadth of the method on four superconducting-circuit tasks: preparation of a five-qubit GHZ state, a CNOT gate, qubit readout, and an error correction primitive, modeled with realistic multilevel transmons, decay, and strong drives, in each case reaching fidelities consistent with the intrinsic dissipation limits. LROC thereby extends pulse-level optimization to system sizes beyond the reach of existing gradient-based methods.

\end{abstract}

\maketitle


\section{\label{sec:introduction}Introduction}

Quantum computers process information by steering controllable quantum systems through sequences of state preparations, gates, and measurements. In the abstract circuit model, these operations are ideal unitaries and projective measurements; in the laboratory, they are implemented by time-dependent control fields acting on an open quantum system. The quality of a processor is therefore limited not only by the target circuit, but by how accurately these pulse-level operations are realized. This is particularly evident in superconducting circuits, where the drives that implement logical operations also excite leakage levels, activate always-on couplings, and compete with relaxation, dephasing, and measurement-induced dissipation~\cite{blais_circuit_2021, krantz_quantum_2019}. As devices grow, hand-designed pulses and sequential calibration become increasingly difficult to scale.

Quantum optimal control provides an alternative: rather than assembling a protocol from idealized gates, one directly optimizes the available control fields against the physical objective~\cite{koch_quantum_2022}. Among the most widely used algorithms is Gradient-Ascent Pulse Engineering (GRAPE), which exploits a piecewise-constant pulse parametrization to obtain analytic gradients with respect to all control amplitudes~\cite{khaneja_optimal_2005}. Its central advantage is that the cost of a gradient is comparable to that of simulating the dynamics itself, making high-dimensional pulse optimization practical~\cite{machnes_comparing_2011, goerz_charting_2017, george_minimal_2025}.

For open quantum systems, however, the simulation itself is the bottleneck. A Hilbert space of dimension $\N$ requires $\N^2$ density-matrix degrees of freedom, which quickly becomes prohibitive for multiqubit devices with resonators and leakage levels. The cost is most severe in optimal control, where the dynamics must be evaluated at every iteration. Recent progress incorporating automatic differentiation (AD) into open-system control routines inherits this cost and adds to it: reverse-mode AD stores the computational graph of the entire evolution, a memory overhead that becomes prohibitive at scale~\cite{abdelhafez_gradient-based_2019, gautier_optimal_2025, lu_optimal_2024}.

A number of strategies have been developed to tame the cost of open-system simulation. Quantum trajectory unravelings replace the master equation by an ensemble of stochastic pure-state evolutions~\cite{dalibard_wave_function_1992, molmer_monte_1996}, while variational parametrizations -- tensor networks~\cite{verstraete_matrix_2004, weimer_simulation_2021} and other low-rank representations~\cite{le_bris_low-rank_2013, joubert-doriol_problem-free_2015, mccaul_fast_2021, santos_low-rank_2025, gravina_adaptive_2024} -- restrict the dynamics to a reduced set of degrees of freedom. Common to the latter is the observation that the states arising in practice are far from generic: in quantum computing in particular, protocols are designed to preserve purity, so the statistical ensemble described by the density matrix is dominated by a small number of pure states. In a previous work~\cite{goutte_low-rank_2026}, we exploited this structure to accelerate the evaluation of a control objective for large systems in the context of transmon readout, treating the optimization routine itself as secondary.

In this work, we complete the program: we combine a low-rank representation of the density matrix~\cite{le_bris_low-rank_2013, joubert-doriol_non-stochastic_2014, goutte_low-rank_2026} with the adjoint-state method~\cite{chen_neural_2019, gautier_optimal_2025} and a piecewise-constant pulse parametrization~\cite{khaneja_optimal_2005} into a low-rank optimal control (LROC) algorithm. Each ingredient is well established; their combination yields an analytic gradient at the same reduced cost as the low-rank evolution itself, with a memory footprint set by the stored trajectory alone. Any differentiable objective enters through only two derivatives, leaving the adjoint dynamics fixed. We demonstrate the method on four superconducting-circuit tasks spanning the operational pipeline: preparation of a GHZ state, a cross-resonance CNOT gate, transmon readout, and a parity-check primitive for error correction. The paper is organized as follows. Section~\ref{sec:formalism} presents the formalism and its scaling; Sec.~\ref{sec:applications} presents the applications; Sec.~\ref{sec:discussion} discusses extensions and limitations. Derivations, common loss-function gradients, and numerical details are collected in the appendices.

\section{\label{sec:formalism}Formalism}

We begin by presenting the formalism underlying LROC. The goal is to express the control signal in terms of a set of parameters, and find the optimal parameter values such that dissipative evolution under this control minimizes a given loss function~\cite{khaneja_optimal_2005, doria_optimal_2011, koch_quantum_2022, petruhanov_grape_2023, abdelhafez_gradient-based_2019, gautier_optimal_2025, xie_optimal_2026}. We consider a driven open quantum system whose state at time $t$ is described by the density matrix $\hat\rho(t)$. If we assume that the Hilbert space has dimension $N$, then the density matrix is expressed as a $N\times N$ self-adjoint matrix. Within the assumption that the system interacts with a Markovian environment, the dynamics of the density matrix is governed by the Lindblad master equation ($\hbar = 1$)
\begin{equation}
    \dot{\hat\rho} = -\I [\hat H, \hat\rho] + \sum_s\mathcal{D}[\hat L_s]\hat \rho,
    \label{eq:ME}
\end{equation}
with Hamiltonian
\begin{equation}
    \hat H = \hat H_0 + \sum_k u_k(t) \hat H_k
\end{equation}
and dissipators $\mathcal{D}[\hat L_s]\hat \rho = \hat L_s \hat \rho \hat L_s^\dagger - \{\hat L_s^\dagger \hat L_s, \hat\rho\} / 2$. Here, $\hat H_0$ is the drift Hamiltonian, $\hat H_k$ are the control Hamiltonians, and $u_k(t)$ are the amplitudes of the control fields to be optimized. The index $k$ labels independent control channels, such as different drive quadratures or different physical ports. The dissipative dynamics, meanwhile, is governed by collapse operators $\hat L_s$ which may correspond to, e.g., population decay, dephasing, or exchange with a thermal bath. The numerical solution of Eq. \eqref{eq:ME} typically requires integrating a set of $N^2$ coupled ordinary differential equations.

We consider a loss function of the form
\begin{equation}\label{eq:loss}
     C = \Phi[\hat\rho(T)] + \int_0^T \phi[\hat\rho(t)]\,\diff{t},
\end{equation}
where $T$ is the total time of the evolution and $\Phi$ and $\phi$ are real-valued terminal and running contributions, respectively. The dependence on the controls $u_k(t)$ is implicit, through the trajectory $\hat\rho(t)$ they generate.

Optimizing this loss function typically requires executing gradient descent in the space of all parameters, thus leading to numerically integrating the density matrix many times. Alternatively, the gradient of the loss function can be computed with a backpropagation algorithm~\cite{schmidt_optimal_2011, chen_neural_2019, chen_robust_2025}. Both the evolution and the gradient computation, however, inherit the $\mathcal{O}(N^2)$ scaling of the computational cost of integrating the Master equation~\cite{breuer_theory_2002, shillito_dynamics_2022}.


\subsection{Low-rank approximation}\label{sec:lra}

Instead of propagating the full density matrix, we represent it as a convex mixture of pure states,
\begin{equation}\label{eq:lra}
    \hat \rho = \sum_{i=1}^{\N} \tilde p_i \ket{\psi_i}\bra{\psi_i},
\end{equation}
where $\tilde p_i$ is the population of the pure state $\ket{\psi_i}$, $\tilde p_i \geq \tilde p_{i+1}$, and $\sum_{i=1}^N \tilde p_i = 1$. The low-rank approximation (LRA) consists of truncating the sum to its $M$ largest $\tilde p_i$ terms,
\begin{equation}
    \hat\rho \simeq \sum_{i = 1}^M p_i \ket{\psi_i}\bra{\psi_i} = \m \m^\dagger,
\end{equation}
where we have introduced the compact representation in terms of the $\N \times M$ matrix $\m$, whose $i$th column is $\sqrt{p_i}\ket{\psi_i}$ and $p_i = \tilde p_i / \sum_{i=1}^M \tilde p_i$~\cite{le_bris_low-rank_2013, joubert-doriol_non-stochastic_2014, joubert-doriol_problem-free_2015, santos_low-rank_2025, goutte_low-rank_2026}. For instance, for a pure state $\hat\rho = \ket{\psi_0}\bra{\psi_0}$, the first column of $\m$ is $\ket{\psi_0}$ and all other columns are zero~\footnote{In practice, all zero columns are populated by a random orthonormal basis of weight $\epsilon$ (see Appendix~\ref{app:implementation}).}.


The low-rank master equation~\cite{joubert-doriol_non-stochastic_2014, goutte_low-rank_2026} then directly governs the evolution of $\m$:
\begin{equation}\label{eq:nosse}
    \dot{\m} = - \I \hat H \m + \mathcal O[\m] - \mu \m = \mathbf f(\m).
\end{equation}
The unitary term $-\I \hat H \m$ is linear and does not mix the columns of $\m$. The dissipative contribution, on the other hand, is
\begin{equation}\label{eq:nosse_dissipation}
    \mathcal O [\m] = \frac{1}{2} \sum_s \left[\hat L_s \m (\m^{+} \hat L_s \m)^\dagger - \hat L_s^\dagger \hat L_s \m \right],
\end{equation}
where $\m^+ = (\m^\dagger \m)^{-1} \m^\dagger$ is the Moore-Penrose pseudo-inverse. The dissipative dynamics generally increases the mixedness through this non-linear term. This growth can be tracked by dynamically adapting $M$~\cite{le_bris_adaptive_2015, joubert-doriol_problem-free_2015,gravina_adaptive_2024}, or absorbed by a suitably large constant $M$ and subsequently verifying convergence. In this work, we employ the latter strategy. Finally, $\mu = \mathrm{Re} \, \mathrm{tr} (\mathcal{O}[\m]^\dagger \m) / \mathrm{tr}(\m^\dagger \m)$ ensures trace-preservation.

Assuming the Hamiltonian and collapse operators are sparse, the numerical integration of Eq.~\eqref{eq:nosse} has computational complexity that scales as $\mathcal{O}(\N M^2)$ in time and $\mathcal O (\N M)$ in memory. Integrating the Lindblad master equation instead scales as $\mathcal{O}(\N^2)$ in both~\cite{weimer_simulation_2021, campaioli_quantum_2024}. When $M \ll \N$, as is the case for most standard quantum computing setups where the purity remains high by design, the LRA dramatically reduces the number of dynamical variables while retaining the essential physics. $M$ therefore controls the strength of the approximation, trading off expressivity for a lighter computational overhead. 

We note that the factorization Eq.~\eqref{eq:lra} is structurally related to other purified or compressed representations, such as locally purified density operators and tensor-network ansätze~\cite{verstraete_matrix_2004, werner_positive_2016, weimer_simulation_2021, guo_locally_2024}. The present form is distinguished by yielding a closed equation of motion for $\m$, Eq.~\eqref{eq:nosse}, which renders it directly amenable to the adjoint-state treatment that follows.

\subsection{Optimization algorithm}\label{sec:optimization}

\begin{figure}[t]
    \centering
    \includegraphics[width=\linewidth]{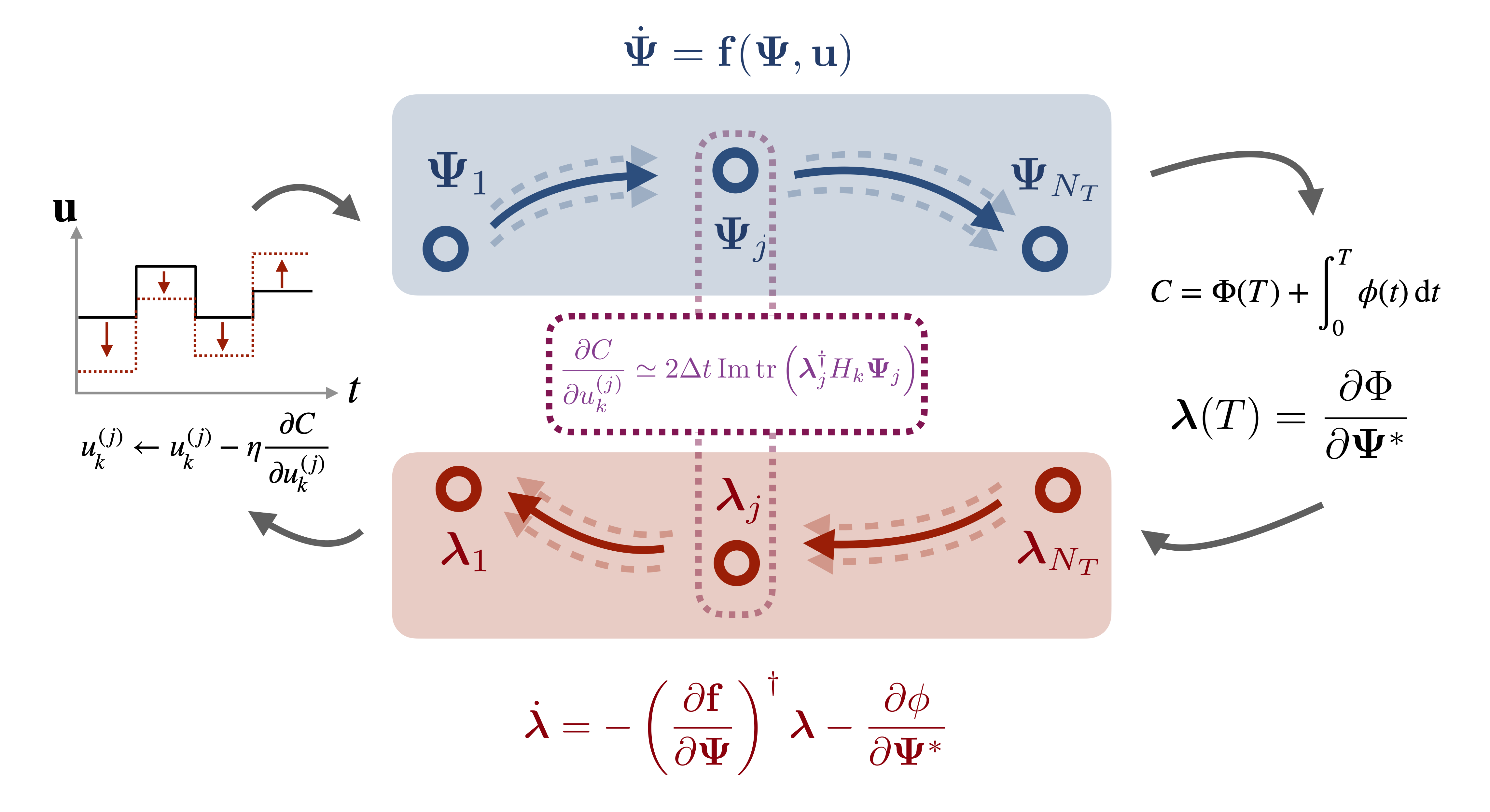}
    \caption{Schematic of the LROC algorithm. A forward pass is evaluated with a candidate pulse and the states $\{\mj\}$ are stored in memory (blue block). From these, the loss is computed and the adjoint state is initialized and $\w (T)$ evolves backward in time following Eq.~\eqref{eq:adjoint_eom} (red block), computing the gradient at every time $t_j = (j-1) \Delta t$ and subsequently discarding the forward state $\m_j$. The parameters are updated through gradient descent or Adam~\cite{kingma_adam_2015} and the process is repeated until convergence. A similar procedure holds for multi-state objectives by parallelizing over many states (faded dashed arrows). Further details on the numerical implementation are provided in Appendix~\ref{app:implementation}.}
    \label{fig:schematic}
\end{figure}

In the context of optimal control, where the loss generally depends on the dynamics of the state, the LRA leads to a speedup in the loss function evaluation~\cite{goutte_low-rank_2026}. In this section, we expand on this idea by incorporating it within a broader pulse optimization routine. In particular, we derive an analytic expression for the loss gradient by combining the LRA with the adjoint state method~\cite{chen_neural_2019, boscain_introduction_2021} and a simple yet general pulse parametrization~\cite{khaneja_optimal_2005, petruhanov_grape_2023, goerz_grapejl_2025, chen_robust_2025}.

We begin with the latter, dividing $T$ into $\NT$ slices of duration $\Delta t=T/\NT$ and using a piecewise-constant parametrization $u_k(t) \to \ukj$, with $j=1,\dots,\NT$. The optimization variables are therefore the finite set of amplitudes $\ukj$. This is identical to the parametrization employed by GRAPE algorithms~\cite{khaneja_optimal_2005}. Within this parametrization, a simple formula for the gradient with respect to the piecewise-constant controls is derived (see Appendix~\ref{app:formalism}),
\begin{align}
    \frac{\partial C}{\partial \ukj} = 2 \Delta t\, \Im\, \trexpval{\wj}{\hat H_k \mj} + \mathcal{O}(\Delta t^2). \label{eq:grape_gradient}
\end{align}
Here we have introduced the adjoint state $\w$, which we define below. 

Equation~\eqref{eq:grape_gradient} deserves a more detailed discussion. This gradient coincides with the first-order GRAPE gradient for a closed system, i.e. when $\m$ and $\bm\lambda$ are pure states of rank one. In this case the adjoint state is evolved backward in time from $\w (T)$ following $\dot{\w} = -\I \hat H \w$~\cite{khaneja_optimal_2005}. An open quantum system described by the Lindblad master equation is treated in a similar way~\cite{petruhanov_grape_2023, chen_robust_2025}. Within the present low-rank evolution Eq.~\eqref{eq:nosse}, an analytical expression for the so-called backward pass may still be derived. In particular,
\begin{align}\label{eq:adjoint_eom}
    \dot{\w} = -\left(\frac{\partial \mathbf f}{\partial \m}\right)^\dagger \w - \frac{\partial \phi}{\partial \m^*}
\end{align}
with terminal condition
\begin{align}
    \w(T) = \frac{\partial \Phi}{\partial \m^*}.
\end{align}
Once the forward states $\mj$ and adjoint states $\wj$ have all been computed, the pulse gradients follow directly from Eq.~\eqref{eq:grape_gradient} and the parameters may be updated via a simple gradient descent step. A schematic of the LROC algorithm is shown in Figure~\ref{fig:schematic}.

Crucially, any differentiable loss enters only through the terminal derivative $\partial\Phi/\partial\m^*$, which fixes the adjoint boundary $\bm\lambda(T)$, and the running derivative $\partial\phi/\partial\m^*$, which enters as a source term in Eq.~\eqref{eq:adjoint_eom}~\footnote{Loss function contributions which depend explicitly on the controls $\ukj$ (e.g. discontinuity and amplitude penalties) and do not depend on $\m$ bypass the adjoint entirely and contribute directly to the gradient, modifying its form. In practice, however, they are enforced through pulse constraints given directly to the optimizer.}. The main part of the adjoint dynamics, $(\partial \mathbf f / \partial \m)^\dagger$, and the gradient computation remain unchanged. Terminal state and gate fidelities, trajectory-dependent running costs, and nonlinear functions thereof can all be accommodated in this way. Explicit expressions for typical loss function contributions, including those used in this work, are collected in Appendix~\ref{app:losses} and a series of applications will follow in Sec.~\ref{sec:applications}.

\subsection{Scaling and advantage}\label{sec:scaling}

\begin{figure}
    \centering
    \includegraphics[width=\linewidth]{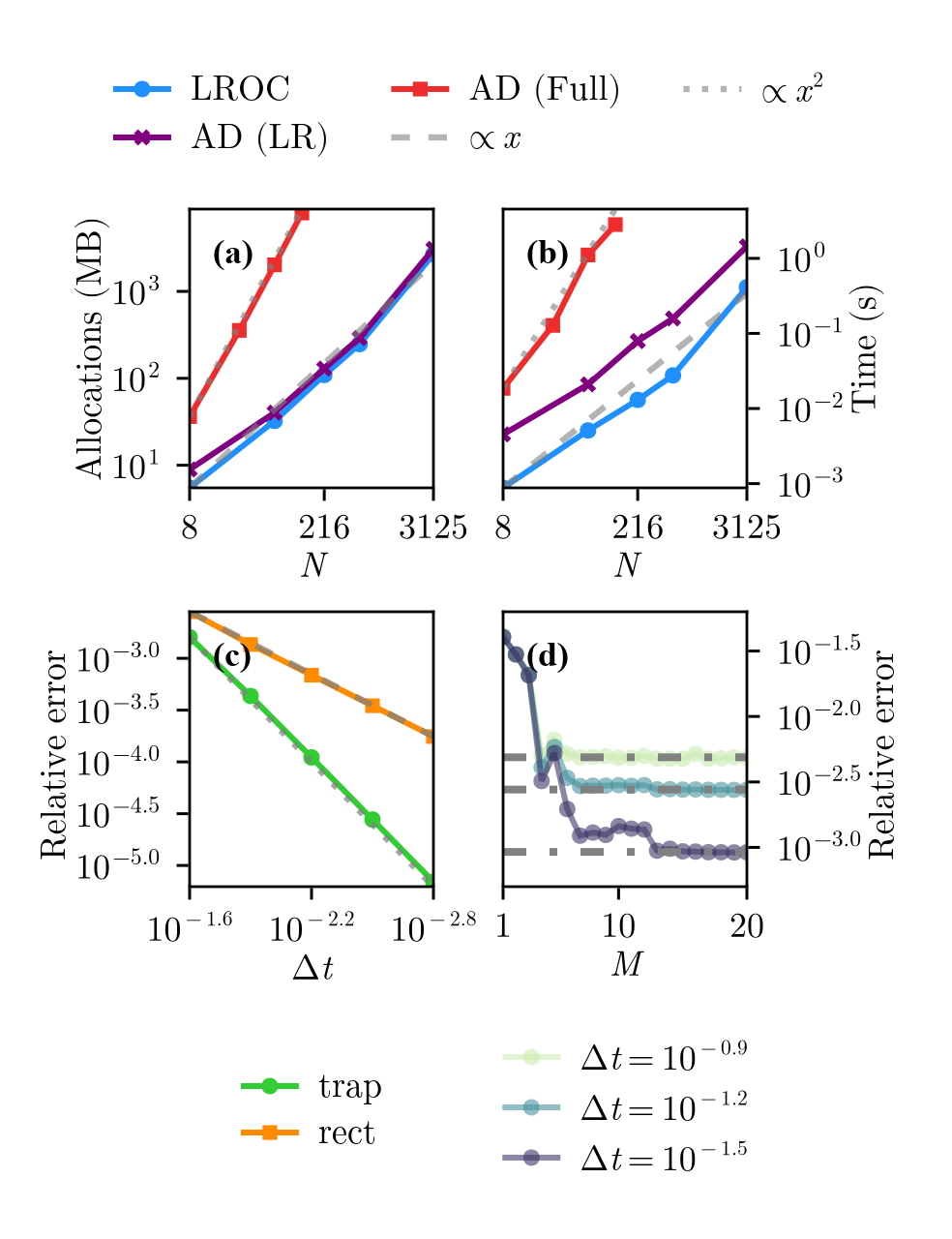}
    \caption{(a) Memory footprint and (b) total runtime as a function of the Hilbert space dimension $N$ for LROC (blue), AD with low-rank (purple), and AD with the full master equation (red). (c) The accuracy of the gradient is controlled by $\Delta t$ and (d) $M$, with the latter hitting the plateau given by the trapezoidal error (dash-dotted gray lines). All plots are shown for a three-qubit GHZ state (see Sec.~\ref{sec:state_fidelity}) where $\N$ is increased by increasing the Hilbert space dimension of the individual multilevel transmons and $M = 2$ [except for (d)]. The AD results were obtained with the \texttt{Enzyme.jl}~\cite{moses_instead_2020, moses_reverse-mode_2021} package.}
    \label{fig:memory_scaling}
\end{figure}

The LROC algorithm consists of three layers: the LRA, the adjoint state method, and a GRAPE-like parametrization. Each contributes its own savings and controlled trade-offs. The first is the LRA, which underlies the decisive advantage of the method by providing a quadratic improvement in both runtime and memory over the Lindblad master equation. Figure~\ref{fig:memory_scaling}(a,b) shows the memory and runtime scaling with $\N$ for a single gradient evaluation in LROC. The results are compared to reverse-mode automatic differentiation (AD) with both low-rank and the full Lindblad master equation. 

The second layer is the adjoint-state method, which makes the gradient computation efficient at the cost of an additional reverse-time evolution. Because the backward pass is an explicit equation [Eq.~\eqref{eq:adjoint_eom}], it stores only the forward trajectory rather than a full computational graph. This differentiates LROC from AD: reverse-mode AD retains every integration substep $n_{\rm sub}$ together with the propagators and their derivatives, so its memory footprint carries an additional factor of $n_{\rm sub}$. In the case of, e.g., rapidly oscillating systems where the adaptive timestep becomes small and $n_{\rm sub}$ grows, this greatly increases the memory overhead of AD. 
Moreover, the pseudo-inverse $\m^+$, which is numerically unstable when $M > \mathrm{rank} \, \left(\m \m^\dagger\right)$, is supplied in closed form, whereas AD must differentiate it directly. A detailed discussion comparing the two methods is supplied in Appendix~\ref{app:comp_ad}.

Nonetheless, for the illustrative example taken in Figure~\ref{fig:memory_scaling}, the scaling of LROC is comparable to that of reverse-mode AD through the low-rank dynamics. The modest improvement comes from the third layer: a GRAPE-like parametrization. This allows for the analytic gradient Eq.~\eqref{eq:grape_gradient} and removes the substep factor $n_{\rm sub}$ (the number of integrator steps within a time slice $\Delta t$) from the gradient computation. Its trade-off is accuracy, as the gradient now carries a quadrature error controlled by $\Delta t$ and a truncation error controlled by $M$ [Figure~\ref{fig:memory_scaling}(c,d)]. The former could be reduced with a higher-order rule, an adaptive integrator, or the augmented-state method~\cite{chen_neural_2019} while the latter can be reduced by increasing $M$.

 \section{Applications}\label{sec:applications}

We test LROC on several applications covering the full range of quantum protocols: state preparation, quantum gates, qubit measurement, and error correction subroutines. 
We focus on superconducting circuits, though we stress that LROC may be applied to any platform provided the purity remains high. 

Throughout the following and unless otherwise stated, we model the qubits as nonlinear oscillators~\cite{blais_circuit_2021}. The Hamiltonian of the $k$th qubit is given by
\begin{equation}\label{eq:transmon}
    \hat H_{k} = \omega_{k} \hat b_k^\dagger \hat b_k - \frac{\alpha}{2} \hat b_k^\dagger \hat b_k^\dagger \hat b_k \hat b_k,
\end{equation}
where $\hat b_k$ is the annihilation operator of qubit $k$, $\omega_k$ is its frequency, and $\alpha$ quantifies the strength of the nonlinearity which we take to be the same for all qubits. This model accurately describes the physics of transmon qubits~\cite{blais_circuit_2021}, where the cosine potential can be expanded up to the $4$-th order term, and holds under the assumption that leakage is restricted to the lowest-lying levels. For the present analysis, we restrict the Hilbert space to the $5$ lowest-lying energy levels. We further assume that a decay process is induced by the environment, assumed to be at zero temperature, with decay rate $\gamma = 1/T_1$ and $T_1 \simeq 15\,\mu\mathrm{s}$. The decay process is accounted for by the collapse operator $\hat L_k = \sqrt{\gamma}\, \hat b_k$ in the Lindblad master equation. The retained non-computational levels and finite lifetime are in line with typical transmon devices~\cite{swiadek_enhancing_2024, arute_quantum_2019}. Moreover, we retain all counter-rotating terms in the Hamiltonian and do not apply the Rotating-Wave Approximation (RWA), as it has been shown to introduce important errors in strongly driven systems~\cite{sank_measurement-induced_2016, kohler_dispersive_2018, khezri_measurement-induced_2023, dumas_measurement-induced_2024, ferrari_dissipative_2025}. All final optimization results are verified with full master-equation simulations via the \texttt{QuantumToolbox.jl} package~\cite{mercurio_quantum_2025}.

Each application herein highlights a different feature of LROC. For instance, the state preparation is a single terminal loss function contribution, while gate fidelity is a multi-state objective and thus parallelizes over many such costs. Qubit measurement introduces a nonlinear running cost, and error correction a nonlinear terminal cost which is also parallelized.

\subsection{\label{sec:state_fidelity}State preparation}

\begin{figure*}
    \centering
    \includegraphics[width=\linewidth]{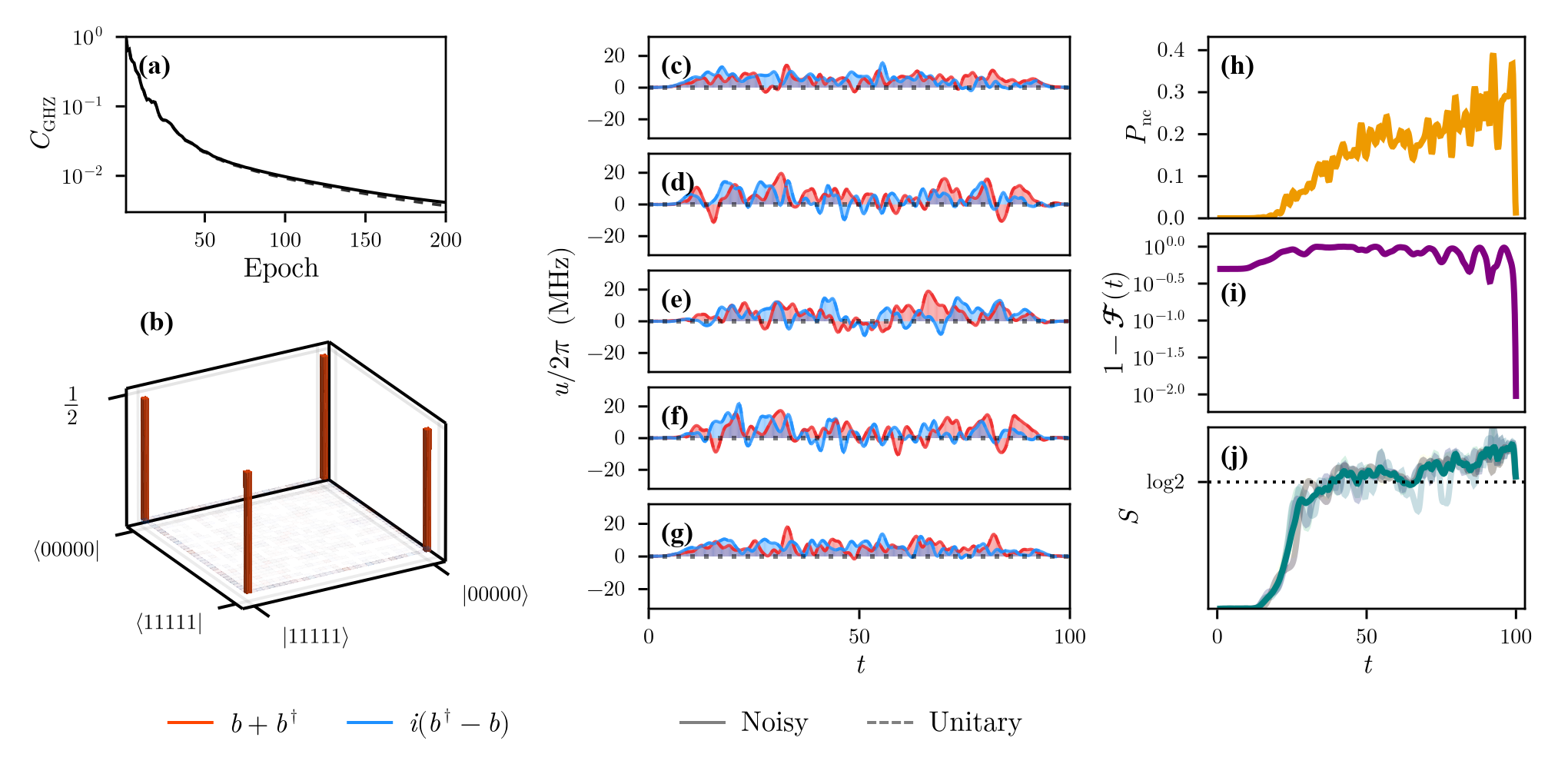}
    \caption{Preparation of a 5-qubit GHZ state. 
    (a) Evolution of the loss function~\eqref{eq:ghz_loss} as a function of the optimization epoch. 
    (b) Final GHZ state tomography. 
    (c)--(g) Optimal control pulses subject to smoothing and clipping constraints (see Appendix~\ref{app:constraints}). 
    (h)--(j): Dynamics of the optimized pulse: (h) Population in the non-computational subspace $P_{\rm nc}$;
    (i) Evolution of the infidelity $1 - \mathcal{F} = C_{\rm GHZ}$;
    (j) von Neumann entanglement entropy $-\mathrm{tr}(\hat \rho_k \log\hat\rho_k)$ of each qubit's reduced density matrix $\hat\rho_k = \mathrm{tr}_{\setminus k} (\hat\rho)$ (faded lines) and their average (bold line).
    Parameters: $M = 2$, $J/2\pi = 50$ MHz, $\omega/2\pi = 5$ GHz, $\alpha/2\pi = 300$ MHz, $\gamma/2\pi = 10$ kHz, $T = 100$ ns.}
    \label{fig:ghz_result}
\end{figure*}

As a first application, we consider the preparation of the $n_{\rm GHZ}$-qubit Greenberger--Horne--Zeilinger (GHZ) state,
\begin{equation}
    \ket{\Psi_{\mathrm{GHZ}}} = \frac{\ket{0}^{\otimes n_{\rm GHZ}} + \ket{1}^{\otimes n_{\rm GHZ}}}{\sqrt{2}}.
\end{equation}
In a circuit model, a GHZ state can be generated from an initial product state by applying a Hadamard gate to one qubit followed by a sequence of CNOT gates that spreads the resulting superposition across the register~\cite{krantz_quantum_2019}. Here, instead of decomposing the operation into gates, we directly optimize a continuous control pulse schedule, whose objective is:
\begin{equation}\label{eq:ghz_loss}
    C_{\mathrm{GHZ}} = 1 - \bra{\Psi_{\mathrm{GHZ}}}\m(T) \m^\dagger (T)\ket{\Psi_{\mathrm{GHZ}}}.
\end{equation}

The Hamiltonian in the lab frame is
\begin{equation}\label{eq:ghz_hamiltonian}
    \hat H_{0, \rm{GHZ}} = \sum_{k=1}^{n_{\rm GHZ}} \hat H_k - \sum_{k=1}^{n_{\rm GHZ} - 1} J\left(\hat b_k^\dagger - \hat b_k\right)\left(\hat b_{k+1}^\dagger - \hat b_{k+1}\right),
\end{equation}
where $J$ is the strength of nearest-neighbor capacitive coupling. The controls are given by
\begin{align}\label{eq:ghz_controls}
\begin{split}
    H_{d, \mathrm{GHZ}} = \sum_{k = 1}^{n_{\rm GHZ}} \Big[ u_{k}^I(t) (\hat b_k^\dagger + \hat b_k) + \I u_{k}^Q(t) (\hat b_k^\dagger - \hat b_k)\Big] \cos(\omega_d t),
\end{split}
\end{align}
where $u_k^{I/Q}(t)$ are the microwave control envelopes to optimize and $\omega_d$ is the frequency of the drive. Note that we retain all counter-rotating terms in the Hamiltonian and do not apply the Rotating-Wave Approximation (RWA), as it has been shown to introduce important errors in strongly driven systems~\cite{sank_measurement-induced_2016, kohler_dispersive_2018, khezri_measurement-induced_2023, dumas_measurement-induced_2024, ferrari_dissipative_2025, goutte_low-rank_2026}.

For the results shown below, the qubits are taken to be resonant, $\omega_k = \omega_d \ $ for all $ \ k$, and the coupling $J$ is fixed. 
Figure~\ref{fig:ghz_result}(b-g) shows the optimized pulse and resulting final-state for $n_{\rm GHZ} = 5$. For $T = 100$ ns, we obtain a final state fidelity $\mathcal F = 99.6 \%$, well above experimentally reported GHZ fidelities in superconducting platforms~\cite{zhu_cross-platform_2022}. The residual infidelity is comparable to that expected from the intrinsic dissipation over $T$.

Interestingly, the optimizer routes the state preparation through the non-computational subspace,
\begin{equation}
    P_{\rm nc} = 1 - \expval{\otimes_{k=1}^5 \hat P_{\rm comp}^{(k)}},
\end{equation}
with $\hat P_{\rm comp}^{(k)} = \ket{0}\bra{0} + \ket{1}\bra{1}$, which is heavily populated before the pulse converges to the target [Figure~\ref{fig:ghz_result}(h)]. The fidelity with the GHZ state and the single-qubit entropy [Figure~\ref{fig:ghz_result}(i,j)] likewise converge sharply towards their expected values of $0$ and $\log 2$, respectively, only at the final time. This first application thus demonstrates the optimizer's ability not only to accommodate the system's imperfections (leakage, decay, counter-rotating terms, and pulse constraints) but indeed to exploit them, all while $\hat\rho$ remains low-rank.


\subsection{\label{sec:gate_fidelity}Average gate fidelity}

\begin{figure*}
    \centering
    \includegraphics[width=\linewidth]{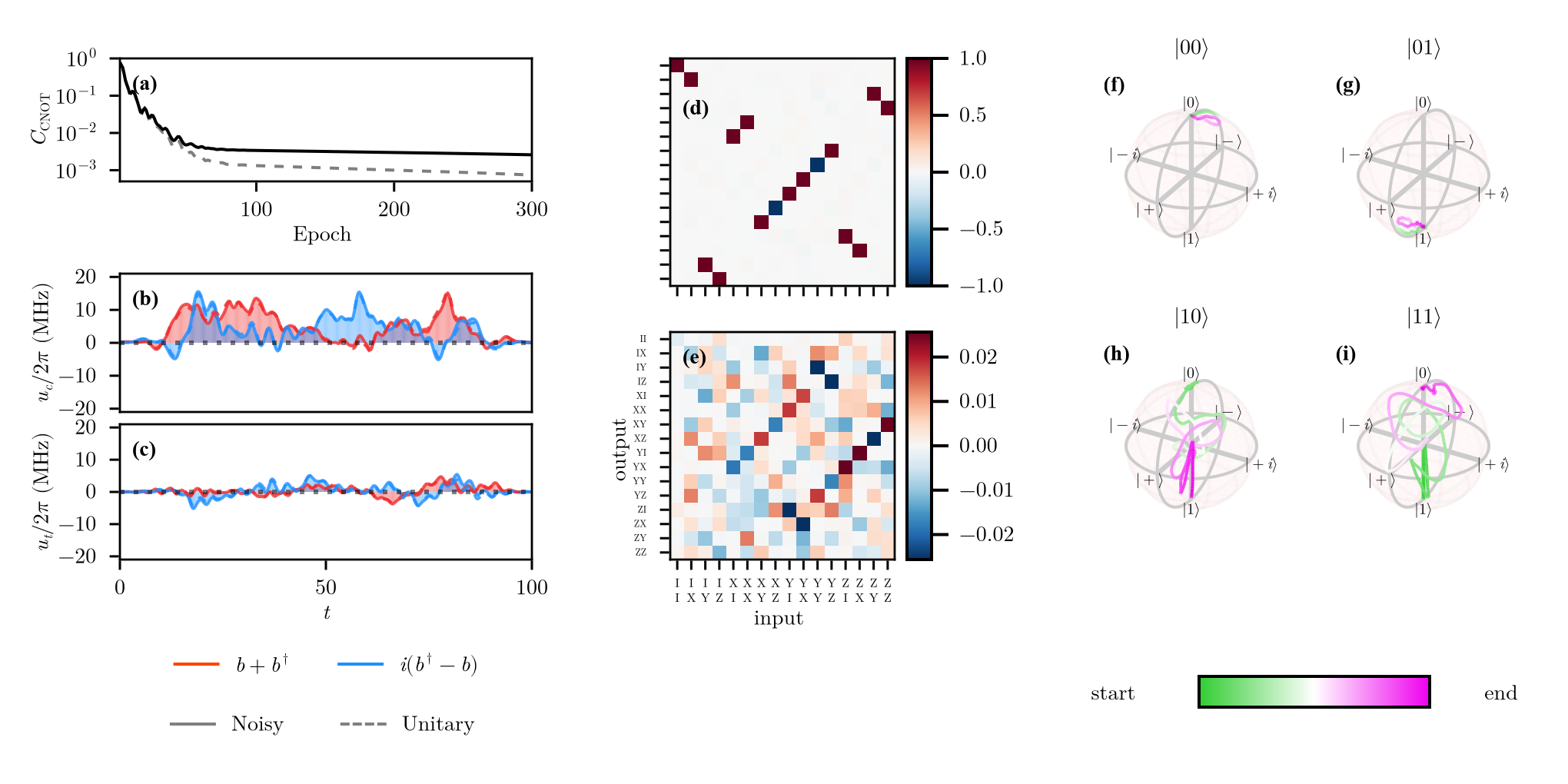}
    \caption{Optimization of a
    the cross-resonance CNOT gate. (a) Loss function~\eqref{eq:cnot_loss} during training. 
    (b) Optimized pulses on the control and (c) target qubits. The unitary pulses are overlaid and match quite closely, though they yield an error one order of magnitude lower. 
    (d) Pauli transfer matrices and (e) error relative to the ideal CNOT gate.
    (f)--(i) Path of the target qubit on its Bloch sphere for each computational basis state under the optimized pulse evolution.
    Parameters: $M = 2$, $J/2\pi = 20$ MHz, $\alpha = 300$ MHz, $\omega_c / 2\pi = 5$ GHz, $\omega_t / 2\pi = 4.75$ GHz, $\gamma / 2\pi = 10$ kHz, $T = 100$ ns.}
    \label{fig:grape_cross_resonance}
\end{figure*}

We now consider the optimization of an entangling two-qubit gate, focusing on the cross-resonance CNOT gate~\cite{rigetti_fully_2010, chow_simple_2011, tripathi_operation_2019, magesan_effective_2020}. The average fidelity between an ideal unitary gate $\hat U$ and its noisy implementation $\mathcal E$ is formally defined as~\cite{nielsen_simple_2002}
\begin{equation}\label{eq:avgfid}
    \bar{\mathcal{F}}({\hat U, \mathcal E}) = \int_\psi d\psi \ \bra{\psi} \hat U^\dagger \mathcal E(\ket{\psi}\bra{\psi}) \hat U\ket{\psi}.
\end{equation}
The integral is over the uniform (Haar) measure in state space, normalized so that $\int d\psi = 1$. This integration is computationally impractical. Here, we instead re-cast Eq.~\eqref{eq:avgfid} as a discrete sum over the states $\ket{\phi_k}$ constituting a 2-design~\cite{dankert_exact_2009, matteo_short_2014}. The loss function is now
\begin{equation}\label{eq:cnot_loss}
    C_{\rm CNOT} = 1 - \frac{1}{K}\sum_{k = 1}^{K} \bra{\phi_k} \hat U^\dagger \mathcal{E}(\ket{\phi_k}\bra{\phi_k}) \hat U\ket{\phi_k},
\end{equation}
where $K = d(d+1)$ and $d$ is the dimension of the logical Hilbert space. For the two-qubit CNOT gate, $d = 4$ and $\{\ket{\phi_k}\}$ are the common eigenstates of five maximally commuting sets of two-qubit Pauli operators~\cite{klimov_geometrical_2007, klimov_discrete_2009}.

The optimization therefore consists of forward-evolving $20$ input states with the same candidate pulse. The states are then combined to compute a single loss function and $20$ adjoint state initializations. The adjoint states are finally backward-evolved to compute the gradient. Both the forward and backward evolutions can be done in parallel, as shown by the faded arrows in Figure~\ref{fig:schematic}.

We focus on a system of a control and a target qubit with frequencies $\omega_{\rm c}$ and $\omega_{\rm t}$, respectively. The Hamiltonian is given by
\begin{equation}\label{eq:cr_hamiltonian}
    \hat H_{0, \rm CNOT} = \hat H_{\rm c} + \hat H_{\rm t} - J(\hat b_\mathrm{c}^\dagger - \hat b_\mathrm{c})(\hat b_\mathrm{t}^\dagger - \hat b_\mathrm{t}),
\end{equation}
where $J$ is the coupling strength and $\hat H_{\rm c,t}$ are the usual transmon Hamiltonians~\eqref{eq:transmon} for the control and target qubit. The controls are analogous to Eq.~\eqref{eq:ghz_controls}. In such fixed-frequency transmon devices, entanglement can be generated by driving the control qubit at the target qubit's dressed frequency, $\omega_d \simeq \omega_{\rm t} - J^2 / \Delta$. Provided the detuning satisfies $|\Delta| = |\omega_\mathrm{c} - \omega_\mathrm{t}| \gg J$, this produces a $ZX$ interaction~\cite{chow_simple_2011, blais_circuit_2021}.

 Figure~\ref{fig:grape_cross_resonance} shows the resulting optimization. The loss function converges to an average gate fidelity of $\mathcal F = 99.6 \%$. 
This value is consistent with the level of performance expected from cross-resonance gates: early demonstrations of the cross-resonance gate reported substantially lower fidelities~\cite{chow_simple_2011}, while subsequent calibration and pulse-shaping techniques pushed two-qubit cross-resonance gate fidelities above $99\%$~\cite{sheldon_procedure_2016}. The present result shows that LROC, like other pulse-shaping techniques~\cite{caneva_chopped_2011, baum_experimental_2021}, can automatically find a high-fidelity shaped pulse without explicitly constructing an echoed cross-resonance sequence, separately calibrating the intermediate $ZX$, or driving precisely at the target qubit's dressed frequency.

Finally, it is worth noting that although the small Hilbert space dimension $N_{\rm CNOT} = 5^2$ in this example does not strictly require a low-rank approach, this application remains an important benchmark demonstrating the natural parallelization of LROC. The present method can be straightforwardly extended to optimize a sequence of gates on a larger circuit or to perform robust optimal control, both of which involve simulating a system with a much higher number of degrees of freedom, making the advantage brought by LROC useful~\cite{george_minimal_2025}. 

\subsection{\label{sec:readout_fidelity}Qubit readout}

\begin{figure*}
    \centering
    \includegraphics[width=\linewidth]{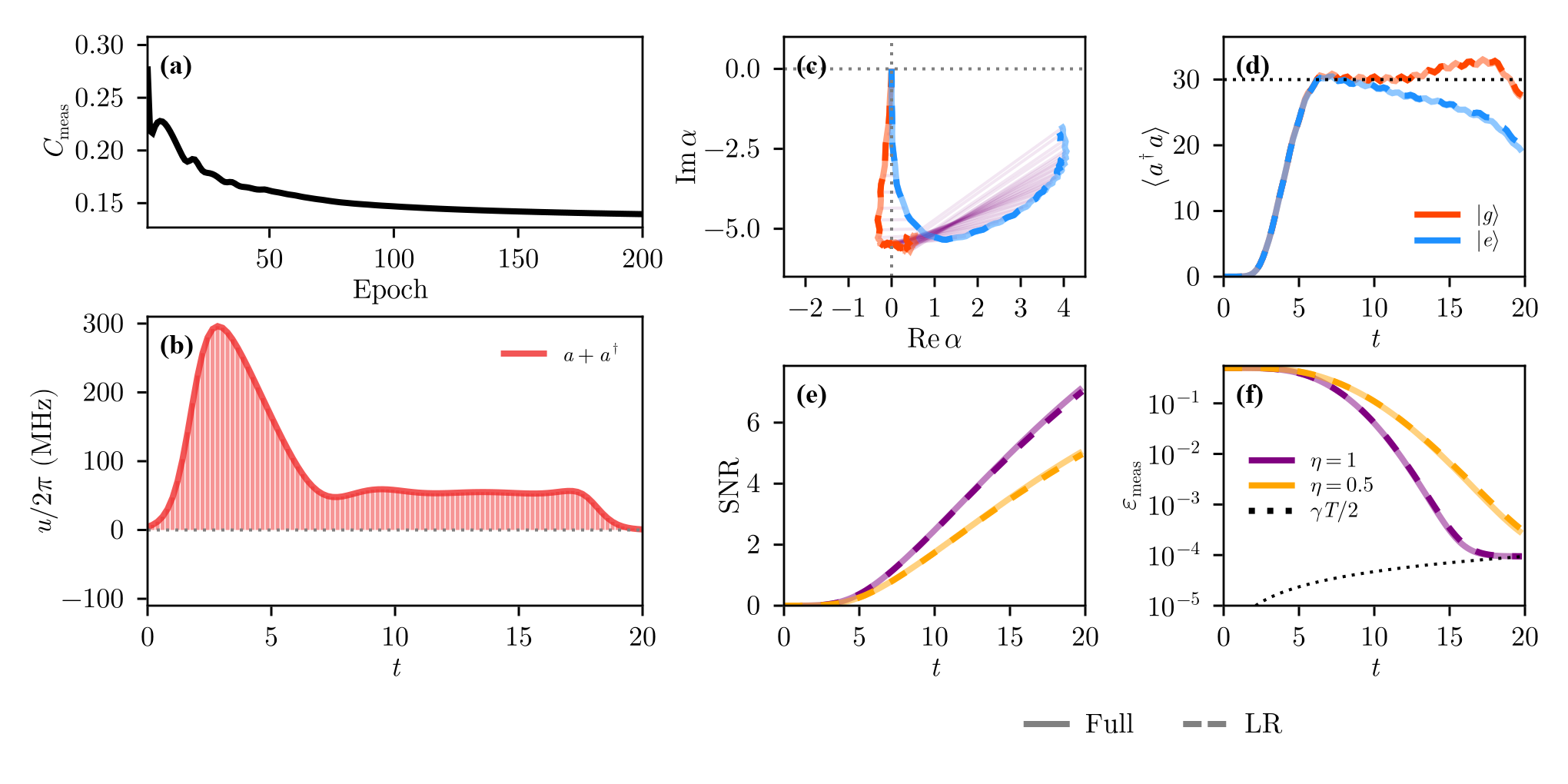}
    \caption{Optimal control of the balanced cross-Kerr transmon readout. (a) Convergence of the loss function Eq.~\eqref{eq:readout_loss} and (b) optimized two-step pulse of the sole . 
    (c) Phase-space trajectories and (d) photon number in the resonator computed with the full (faded lines) and low-rank (dashed lines) dynamics. The loss incurs a penalty when $\expval{a^\dagger a} > n_{\rm max} = 30$ (dotted black line) to stave off ionization~\cite{cohen_reminiscence_2022, shillito_dynamics_2022, dumas_measurement-induced_2024}. 
    (e) Signal-to-noise ratio and (f) assignment error for measurement efficiency $\eta = 1$ (purple lines) and $\eta = 0.5$ (orange lines). 
    Parameters: $E_C / 2\pi = 300$ MHz, $E_{J_c} + E_J = 50 E_C$, $E_{J_C} / 2\pi = 8$ GHz, $\omega_\mathrm r / 2\pi = 9.375$ GHz, $\gamma/2\pi = 1.5$ kHz, $T = 20$ ns.}
    \label{fig:transmon_result}
\end{figure*}

The fast, high-fidelity readout of a transmon qubit remains a central bottleneck in superconducting processors. It has accordingly been a recurring target of pulse shaping, from hand-designed two-step and reset pulses~\cite{boutin_resonator_2017, gautier_optimal_2025} to reinforcement-learned~\cite{chatterjee_enhanced_2025, genois_quantum_2025} and gradient-based model-based optimization~\cite{bengtsson_model-based_2024, abdelhafez_gradient-based_2019}. These efforts confront a difficulty the preceding applications do not share. Unlike state preparation or gates, where dissipation is a small perturbation to be accounted for, readout \emph{requires} interacting with the environment and carries large dissipation by design.


The large dissipation stems from the measurement architecture. To measure a transmon qubit, one couples it to a linear resonator via either a transversal~\cite{blais_cavity_2004} or longitudinal coupling~\cite{didier_fast_2015}. Upon the application of a strong microwave drive, the resonator's phase reveals a dependence on the initial transmon state. The quality of a readout is then dictated by the separation in phase-space of the resonator state conditioned on the transmon being initialized in the ground state $\ket{g}$ and in the excited state $\ket{e}$. This signal-to-noise ratio is a typical measure of the readout fidelity, and is defined as
\begin{equation}\label{eq:snr}
    \mathrm{SNR} = 2\eta \kappa \int_0^T |\alpha_g(t) - \alpha_e(t)|^2 \, \diff{t}
\end{equation}
where $\alpha_q = \mathrm{tr} (\m_q^\dagger \hat a \m_q)$ ($q = g, e$) is the resonator's phase amplitude, $\hat a$ its annihilation operator, $\kappa$ is its decay rate, and $\eta$ is the measurement efficiency. The assignment error can then be obtained from~\cite{bengtsson_model-based_2024} 
\begin{equation}
    \varepsilon_{\rm meas} = \frac{1}{2}\mathrm{erfc}\left(\frac{\sqrt{\rm SNR}}{2}\right) + \frac{\gamma T}{2},
\end{equation}
where $\mathrm{erfc}(x)$ is the complementary error function.

Recent works~\cite{chapple_balanced_2025, wang_longitudinal_2025, mori_high-power_2025, beaulieu_fast_2026} have demonstrated a native cross-Kerr coupling between the transmon and resonator which admits a large SNR in a short time while reducing the Purcell decay rate and remaining robust against ionization. The Hamiltonian for this so-called \emph{balanced cross-Kerr} readout is given by~\cite{chapple_balanced_2025}
\begin{equation}
    \hat H_{0, \rm meas} = \hat H_{\rm trans} + \omega_{\rm r} \hat a^\dagger \hat a - E_{J_c} \cos{\left(\hat \varphi_{\rm r} - \hat \varphi_{\rm t}\right)} + \I J(\hat a^\dagger - \hat a)\hat n_{\rm t}.
\end{equation}
The transmon Hamiltonian is $\hat H_{\rm trans} = 4E_C \hat n_{\rm t}^2 - E_J \cos{\hat\varphi_{\rm t}}$, beyond the nonlinear oscillator approximation~\eqref{eq:transmon} since higher transmon levels outside the Josephson well must be accounted for. The transmon's phase and charge operators are $\hat \varphi_{\rm t}$ and $\hat n_{\rm t}$, respectively, satisfying $[\hat \varphi_{\rm t}, \hat n_{\rm t}] = i$. The resonator frequency is $\omega_{\rm r}$, $E_{J_c}$ is the strength of the junction coupling, and $J$ the capacitive coupling. The dissipation is governed by the collapse operators $L_{\kappa} = \sqrt{\kappa} \hat a$ and $L_{\gamma} = \sqrt{\gamma} \hat b$. There is only a single drive on the resonator mode, $H_{d, \mathrm{meas}}= u(t) (a  + a^\dagger) \cos{(\omega_d t)}$.

The optimization landscape is defined by two competing objectives: maximizing the SNR in as short a time as possible, and minimizing leakage, ionization, and other spurious effects which reduce the quality of readout. To this end, we devise a simple loss function whose contributions are all (nonlinear and multi-state) running costs\footnote{Because both contributions are running costs, the gradients for this application are evaluated with the augmented state formulation detailed in Appendix~\ref{app:implementation}.}:
\begin{equation}\label{eq:readout_loss}
    C_{\mathrm{meas}} = \frac{c_{\mathrm{SNR}}}{\sqrt{\mathrm{SNR}}} + c_{n_{\rm max}} \sum_{q = g, e}\int_0^T \mathrm{ReLU}(n_{{\rm r}, q} - n_{\rm max}) \, \diff{t},
\end{equation}
where $n_{{\rm r}, q} = \mathrm{tr}(\m_q^\dagger \hat a^\dagger \hat a \m_q)$, $\m_q$ is the low-rank state for the $q$ trajectory, and ${\rm ReLU}(x) = \max(0,x)$ is the rectified linear unit function.

The results of the optimization with $M = 4$ are shown in Figure~\ref{fig:transmon_result}. We recover the expected two-step pulse: a large initial ramp-up to populate the readout cavity followed by a plateau that respects the photon-number constraint $n_{\rm r} < n_{\rm max}$, chosen well below the critical photon number $n_{\rm crit} \simeq 65$ to avoid ionization~\cite{dumas_measurement-induced_2024}. With this $20$ ns pulse and a qubit lifetime $T_1 \simeq 100 \, \mu \mathrm s$, a measurement error of $\varepsilon_{\rm meas} = 9.4 \times 10^{-5}$ is achieved, limited almost entirely by qubit decay ($\gamma T/2 \simeq 9.4 \times 10^{-5}$; the discrimination error itself is $\varepsilon_{\rm meas} - \gamma T / 2 = 2.2 \times 10^{-7}$). Moreover, accounting for a measurement efficiency $\eta = 0.5$ yields an error of $\varepsilon_{\rm meas} = 2.7 \times 10^{-4}$~\footnote{The quoted assignment errors stem from~\eqref{eq:snr} and thus suppose that the resonator is in a coherent state centered at $\alpha_{g,e}$. In balanced cross-Kerr readout, the resonator may exhibit a significant nonlinearity which renders the linear-filter SNR sub-optimal. Nonetheless, this approximation is shown in Ref.~\cite{chapple_balanced_2025} to be in good agreement with full stochastic Schrödinger equation simulations for these circuit parameters, i.e. $\varepsilon_{\rm meas} \simeq [P(g|e) + P(e|g)]/2$.}. 

This method can be straightforwardly extended to include penalties for QNDness~\cite{mori_high-power_2025} and fast resonator reset~\cite{boutin_resonator_2017, geher_reset_2025}, and applies equally to other emerging readout schemes. High-frequency readout~\cite{kurilovich_high-frequency_2025, mencia_raising_2025, dixit_millimeter_2026} is a particularly natural target: it operates in regimes of large photon number and rapidly oscillating dynamics where traditional approximations fail~\cite{dai_spectroscopy_2025} and full master-equation simulation is impractical, which are precisely the conditions under which the low-rank representation retains its advantage. 

\subsection{\label{sec:error_correction}Error correction}

\begin{figure*}
    \centering
    \includegraphics[width=\linewidth]{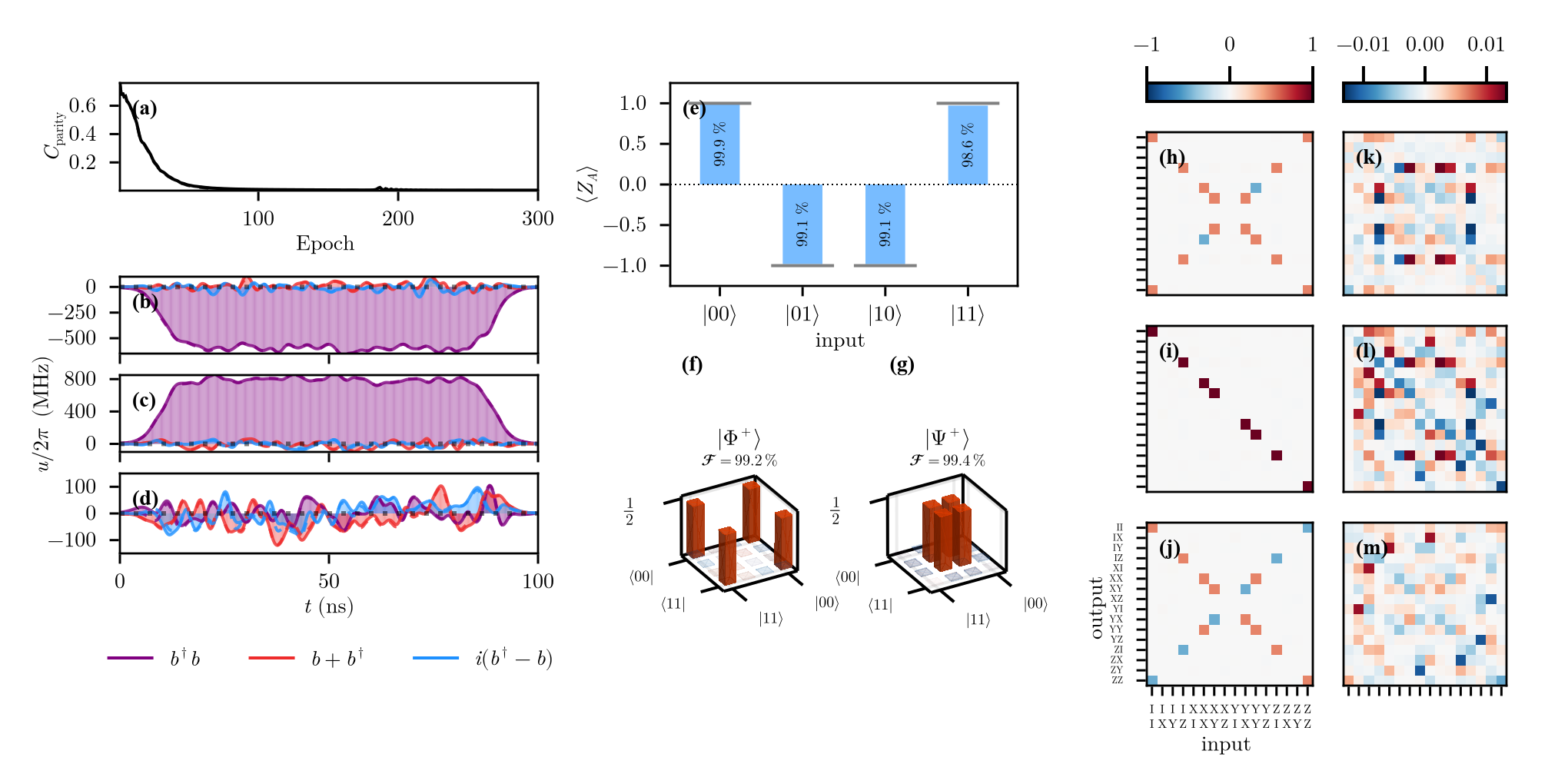}
    \caption{Optimization of the mutual parity measurement. (a) Evolution of the loss function~\eqref{eq:qec_loss}. (b)-(d) Optimal pulses. The frequency of the transmon qubits is directly tuned by the flux pulses (purple) while additional control is supplied by the $I$ (red) and $Q$ (blue) drives (multiplied by $10$ for visibility). (e) Fidelity of the mutual parity measurement in the computational basis. (f) State tomography of the post-selected even and (g) odd Bell states with fidelity $99.2 \, \%$ and $99.4 \, \%$, respectively. 
    (h)--(j) Pauli transfer matrices of (i) unconditional, (h) even-conditioned, (j) and odd-conditioned parity measurement and (k)--(m) the error relative to the ideal measurement.
    Parameters: $M = 2$, $\omega_1 / 2\pi = 5.6$ GHz, $\omega_2 / 2\pi = 4.2$ GHz, $\omega_A /2\pi = 5.0$ GHz, $\alpha / 2\pi = 264$ MHz, $J / 2\pi= 30$ MHz, $\gamma / 2\pi= 10$ kHz, $T = 100$ ns.}
    \label{fig:qec_result}
\end{figure*}

The final application, quantum error correction, combines the ingredients of the preceding two: entangling operations on data qubits steered by a measurement-like objective on an ancilla. Concretely, repetition-based QEC codes are built from mutual parity checks represented by the operator $Z_1 Z_2$. In practice, the value of $Z_1 Z_2$ is encoded in a single ancilla coupled to both data qubits. A series of entangling gates then reveals the value: if $Z_1Z_2 = +1$, the ancilla remains in its initial $\ket{0}$ state, whereas if $Z_1Z_2 = -1$ the ancilla state is flipped to $\ket{1}$, flagging a bit-flip error.

Here, we simulate a single one of these elementary parity measurements. The Hamiltonian consists of two data qubits each coupled to a single ancilla qubit,
\begin{equation}
    \hat H_{0, \rm QEC} = \sum_{k=1,2,A} \hat H_k - \sum_{k = 1, 2} J (\hat b_A^\dagger - \hat b_A)(\hat b_k^\dagger - \hat b_k).
\end{equation}
The entangling mechanism of choice is a flux-tunable controlled-phase gate, in contrast to the cross-resonance gate implemented in Sec.~\ref{sec:gate_fidelity}. The drive operators are therefore the usual $u^{I,Q}$ drives with an additional flux pulse $u^{\rm flux}(t) \hat b^\dagger \hat b$ on each qubit for a total of nine controls.

The figure of merit of a successful mutual parity measurement is the projection onto the $\ket{0}$ ($\ket{1}$) ancilla state for an initial even (odd) physical state~\cite{saira_entanglement_2014, kjaergaard_superconducting_2020}. More precisely, an arbitrary two-qubit state $\ket{\psi_2} = \ket{\psi_2^+} + \ket{\psi_2^-}$, where $\ket{\psi_2^\pm}$ are its (unnormalized) even and odd parts, maps accordingly: 
\begin{equation}\label{eq:qec_target_state}
    \ket{\psi_2} \otimes \ket{0_A} \to \ket{\psi_2^+} \otimes \ket{0_A} + e^{\I  \chi}\ket{\psi_2^-} \otimes \ket{1_A}.
\end{equation}
Note the inclusion of the relative phase $e^{\I \chi}$. Indeed, only the projection onto the ancilla's computational basis is relevant for the parity measurement.
It would therefore be over-constraining to impose $\chi = 0$ by taking a coherence-preserving Haar-averaged state fidelity with target state Eq.~\eqref{eq:qec_target_state} as a loss function. 
Instead, we adopt the loss function
\begin{equation}\label{eq:qec_loss}
    C_{\mathrm{QEC}} = 1 - \frac{1}{K} \sum_{k = 1}^{K} \left( \sqrt{ p_{k}^{+}} + \sqrt{ p_{k}^{-}} \right)^2,
\end{equation}
where
\begin{equation}
    p_{k}^{\pm} = \bra{\Psi_{k}^{\pm}} \m_{k} (T) \m_{k}^\dagger (T) \ket{\Psi_k^{\pm}},
\end{equation}
and
\begin{subequations}\label{eq:qec_even_odd_branches}
    \begin{align}
    \ket{\Psi_k^{+}} =& \ket{\phi_k^+} \otimes \ket{0_A}, \\
    \ket{\Psi_k^{-}} =& \ket{\phi_k^-} \otimes \ket{1_A},
\end{align}
\end{subequations}
are the even and odd branches, respectively, for the physical qubit state $\ket{\phi_k}$ in the $2$-design~\footnote{It should be noted that since Eq.~\eqref{eq:qec_loss} is no longer a second-order polynomial, this choice does not strictly reproduce an average over Haar-distributed states. It remains a good sampling of the $4$-dimensional Hilbert space and we find it to be largely sufficient for our purposes.}. While the loss disregards the coherences of the ancilla qubit, it enforces preservation of the data qubit state $\ket{\psi_2}$. 

The results are shown in Figure~\ref{fig:qec_result} for an optimization with $M = 2$ over a total time of $T = 100$ ns. The average fidelity of mutual parity measurement over the computational basis states~\cite{ali_surface-code_2025} is $99.2 \%$. Over the entire $2$-design, which includes superpositions of even- and odd-parity states, the fidelity of the operation is $99.1 \% $. An often-cited metric for a successful mutual parity check is the pre-measurement fidelity of an initial $\ket{\psi_2} = \ket{++}$ state~\cite{andersen_entanglement_2019, remm_implementing_2023}, which we find to be $99.3 \%$. The fidelity of the ensuing post-selected even and odd Bell states~\cite{marques_logical-qubit_2022, remm_implementing_2023, ali_surface-code_2025} is $99.2  \%$ and $99.4  \%$, respectively. The corresponding Pauli transfer matrices of the $Z$-parity check operation for both pre- and post-selected states are shown~\cite{chow_implementing_2014}, and a maximum error of $1.3 \times 10^{-2}$ is admitted in the $YX$-$IZ$ channel.

\section{Discussion}\label{sec:discussion}

The applications of Sec.~\ref{sec:applications} span the operational pipeline of quantum computing: state preparation, entangling gates, dispersive measurement, and an error-correction primitive. Beyond this practical coverage, they exercise the full physical range for which LROC is suited --- coherent leakage into non-computational levels, relaxation and dephasing, dynamics sampled over the entire Bloch sphere, and large dissipation --- unified by the sole requirement that the state maintain high purity, a natural corollary of most quantum computing protocols. In each case, fidelities consistent with the state-of-the-art were obtained with typical circuit parameters and short times $T \leq 100$ ns; where a dissipation-limited bound exists, as for the CNOT gate of Sec.~\ref{sec:gate_fidelity}, the optimized fidelity approaches it, indicating that the residual infidelity is set by intrinsic dissipation rather than by the optimization.

From a technical perspective, each application is driven by a distinct objective structure: a linear terminal fidelity, a parallelized multi-state average, a nonlinear running cost, and a nonlinear terminal objective. All enter through the final adjoint state ${\partial \Phi}/{\partial \m^*}$ and the source term $\partial\phi/\partial\m^*$ alone. That a single fixed adjoint equation accommodates all of them is the central practical message of this work.

Several extensions follow naturally from this framework. The first is robust control: averaging the objective over a distribution of parameter values reduces the optimization to many forward and backward evolutions propagated under a shared pulse, the same parallelized structure as the multi-state objectives of Secs.~\ref{sec:gate_fidelity} and \ref{sec:error_correction}~\cite{kosut_robust_2013, chen_robust_2025, george_minimal_2025}. The second is the optimization of static parameters themselves, such as couplings or frequencies. Because a static parameter acts at all times, the piecewise-constant locality underlying Eq.~\eqref{eq:grape_gradient} no longer applies; the gradient instead takes the form of a time integral of the adjoint pairing over the full trajectory, accumulated alongside the backward pass via the augmented-state method~\cite{chen_neural_2019} at negligible additional cost.


The method is not without limitations. It is susceptible to local minima, which for weakly dissipative systems could be mitigated by annealing the dissipation rate $\gamma$ and the learning rate over the optimization schedule. In dissipation-dominated regimes, the backward pass instead suffers gradient inaccuracy from the unstable growth of the forward states; this is controlled either by increasing $\NT$ or by evaluating the loss and source terms on a finer sub-grid. A dynamical adaptive-rank scheme, in which $M$ tracks the instantaneous mixedness, would address both the stability and the efficiency of the fixed-$M$ approximation and is a natural direction for future work. Finally, the low-rank factorization compresses mixedness but not entanglement: each column of $\m$ remains a full $\N$-dimensional vector. For registers whose Hilbert-space dimension is itself prohibitive, compressing the columns as tensor networks in a locally purified structure~\cite{werner_positive_2016, guo_locally_2024} is a complementary direction which the closed-form adjoint of Appendix~\ref{app:formalism} can be extended to accommodate.

\begin{acknowledgments}
    We acknowledge support from the Swiss National Science Foundation through Projects No. 200020\_215172, 200021-227992, and 20QU-1\_215928, and as a part of NCCR SPIN (grant number 225153).
\end{acknowledgments}

\section*{Data availability}

To facilitate broader use, our code is openly available at Ref.~\cite{goutte_zenodo_2026}.

\bibliographystyle{apsrev}
\bibliography{references}

@article{blais_circuit_2021,
	title = {Circuit quantum electrodynamics},
	volume = {93},
	issn = {0034-6861, 1539-0756},
	url = {https://link.aps.org/doi/10.1103/RevModPhys.93.025005},
	doi = {10.1103/RevModPhys.93.025005},
	language = {en},
	number = {2},
	urldate = {2024-09-26},
	journal = {Reviews of Modern Physics},
	author = {Blais, Alexandre and Grimsmo, Arne L. and Girvin, S. M. and Wallraff, Andreas},
	month = may,
	year = {2021},
	pages = {025005},
}

@article{gautier_optimal_2025,
  title = {Optimal Control in Large Open Quantum Systems: The Case of Transmon Readout and Reset},
  author = {Gautier, Ronan and Genois, \'Elie and Blais, Alexandre},
  journal = {Phys. Rev. Lett.},
  volume = {134},
  issue = {7},
  pages = {070802},
  numpages = {7},
  year = {2025},
  month = {Feb},
  publisher = {American Physical Society},
  doi = {10.1103/PhysRevLett.134.070802},
  url = {https://link.aps.org/doi/10.1103/PhysRevLett.134.070802}
}

@misc{cohen_reminiscence_2022,
	title = {Reminiscence of classical chaos in driven transmons},
	url = {http://arxiv.org/abs/2207.09361},
	language = {en},
	urldate = {2024-09-26},
	publisher = {arXiv},
	author = {Cohen, Joachim and Petrescu, Alexandru and Shillito, Ross and Blais, Alexandre},
	month = nov,
	year = {2022},
	note = {arXiv:2207.09361 [quant-ph]},
}

@article{dumas_measurement-induced_2024,
	title = {Measurement-{Induced} {Transmon} {Ionization}},
	volume = {14},
	issn = {2160-3308},
	url = {https://link.aps.org/doi/10.1103/PhysRevX.14.041023},
	doi = {10.1103/PhysRevX.14.041023},
	language = {en},
	number = {4},
	urldate = {2024-10-28},
	journal = {Physical Review X},
	author = {Dumas, Marie Frédérique and Groleau-Paré, Benjamin and McDonald, Alexander and Muñoz-Arias, Manuel H. and Lledó, Cristóbal and D’Anjou, Benjamin and Blais, Alexandre},
	month = oct,
	year = {2024},
	pages = {041023},
}

@article{kohler_dispersive_2018,
	title = {Dispersive readout: {Universal} theory beyond the rotating-wave approximation},
	volume = {98},
	issn = {2469-9926, 2469-9934},
	shorttitle = {Dispersive readout},
	url = {https://link.aps.org/doi/10.1103/PhysRevA.98.023849},
	doi = {10.1103/PhysRevA.98.023849},
	language = {en},
	number = {2},
	urldate = {2024-10-31},
	journal = {Physical Review A},
	author = {Kohler, Sigmund},
	month = aug,
	year = {2018},
	pages = {023849},
}

@article{blais_cavity_2004,
	title = {Cavity quantum electrodynamics for superconducting electrical circuits: {An} architecture for quantum computation},
	volume = {69},
	copyright = {http://link.aps.org/licenses/aps-default-license},
	issn = {1050-2947, 1094-1622},
	shorttitle = {Cavity quantum electrodynamics for superconducting electrical circuits},
	url = {https://link.aps.org/doi/10.1103/PhysRevA.69.062320},
	doi = {10.1103/PhysRevA.69.062320},
	language = {en},
	number = {6},
	urldate = {2024-11-05},
	journal = {Physical Review A},
	author = {Blais, Alexandre and Huang, Ren-Shou and Wallraff, Andreas and Girvin, S. M. and Schoelkopf, R. J.},
	month = jun,
	year = {2004},
	pages = {062320},
}

@article{khezri_measurement-induced_2023,
	title = {Measurement-induced state transitions in a superconducting qubit: {Within} the rotating-wave approximation},
	volume = {20},
	issn = {2331-7019},
	shorttitle = {Measurement-induced state transitions in a superconducting qubit},
	url = {https://link.aps.org/doi/10.1103/PhysRevApplied.20.054008},
	doi = {10.1103/PhysRevApplied.20.054008},
	language = {en},
	number = {5},
	urldate = {2024-11-05},
	journal = {Physical Review Applied},
	author = {Khezri, Mostafa and Opremcak, Alex and Chen, Zijun and Miao, Kevin C. and McEwen, Matt and Bengtsson, Andreas and White, Theodore and Naaman, Ofer and Sank, Daniel and Korotkov, Alexander N. and Chen, Yu and Smelyanskiy, Vadim},
	month = nov,
	year = {2023},
	pages = {054008},
}

@article{bengtsson_model-based_2024,
  title = {Model-Based Optimization of Superconducting Qubit Readout},
  author = {Bengtsson, Andreas and Opremcak, Alex and Khezri, Mostafa and Sank, Daniel and Bourassa, Alexandre and Satzinger, Kevin J. and Hong, Sabrina and Erickson, Catherine and Lester, Brian J. and Miao, Kevin C. and Korotkov, Alexander N. and Kelly, Julian and Chen, Zijun and Klimov, Paul V.},
  journal = {Phys. Rev. Lett.},
  volume = {132},
  issue = {10},
  pages = {100603},
  numpages = {6},
  year = {2024},
  month = {Mar},
  publisher = {American Physical Society},
  doi = {10.1103/PhysRevLett.132.100603},
  url = {https://link.aps.org/doi/10.1103/PhysRevLett.132.100603}
}

@article{shillito_dynamics_2022,
  title = {Dynamics of Transmon Ionization},
  author = {Shillito, Ross and Petrescu, Alexandru and Cohen, Joachim and Beall, Jackson and Hauru, Markus and Ganahl, Martin and Lewis, Adam G.M. and Vidal, Guifre and Blais, Alexandre},
  journal = {Phys. Rev. Appl.},
  volume = {18},
  issue = {3},
  pages = {034031},
  numpages = {11},
  year = {2022},
  month = {Sep},
  publisher = {American Physical Society},
  doi = {10.1103/PhysRevApplied.18.034031},
  url = {https://link.aps.org/doi/10.1103/PhysRevApplied.18.034031}
}

@misc{kurilovich_high-frequency_2025,
	title = {High-frequency readout free from transmon multi-excitation resonances},
	url = {http://arxiv.org/abs/2501.09161},
	urldate = {2025-01-17},
	publisher = {arXiv},
	author = {Kurilovich, Pavel D. and Connolly, Thomas and Bøttcher, Charlotte G. L. and Weiss, Daniel K. and Hazra, Sumeru and Joshi, Vidul R. and Ding, Andy Z. and Nho, Heekun and Diamond, Spencer and Kurilovich, Vladislav D. and Dai, Wei and Fatemi, Valla and Frunzio, Luigi and Glazman, Leonid I. and Devoret, Michel H.},
	month = jan,
	year = {2025},
	note = {arXiv:2501.09161},
}

@article{chapple_balanced_2025,
  title = {Balanced Cross-Kerr Coupling for Superconducting Qubit Readout},
  author = {Chapple, Alex A. and Benhayoune-Khadraoui, Othmane and Richer, Simon and Blais, Alexandre},
  journal = {Phys. Rev. Lett.},
  volume = {135},
  issue = {25},
  pages = {256002},
  numpages = {8},
  year = {2025},
  month = {Dec},
  publisher = {American Physical Society},
  doi = {10.1103/r4v5-wyyt},
  url = {https://link.aps.org/doi/10.1103/r4v5-wyyt}
}

@article{swiadek_enhancing_2024,
	title = {Enhancing {Dispersive} {Readout} of {Superconducting} {Qubits} through {Dynamic} {Control} of the {Dispersive} {Shift}: {Experiment} and {Theory}},
	volume = {5},
	issn = {2691-3399},
	shorttitle = {Enhancing {Dispersive} {Readout} of {Superconducting} {Qubits} through {Dynamic} {Control} of the {Dispersive} {Shift}},
	url = {https://link.aps.org/doi/10.1103/PRXQuantum.5.040326},
	doi = {10.1103/PRXQuantum.5.040326},
	language = {en},
	number = {4},
	urldate = {2025-05-09},
	journal = {PRX Quantum},
	author = {Swiadek, François and Shillito, Ross and Magnard, Paul and Remm, Ants and Hellings, Christoph and Lacroix, Nathan and Ficheux, Quentin and Zanuz, Dante Colao and Norris, Graham J. and Blais, Alexandre and Krinner, Sebastian and Wallraff, Andreas},
	month = nov,
	year = {2024},
	pages = {040326},
}

@article{sank_measurement-induced_2016,
	title = {Measurement-{Induced} {State} {Transitions} in a {Superconducting} {Qubit}: {Beyond} the {Rotating} {Wave} {Approximation}},
	volume = {117},
	copyright = {http://link.aps.org/licenses/aps-default-license},
	issn = {0031-9007, 1079-7114},
	shorttitle = {Measurement-{Induced} {State} {Transitions} in a {Superconducting} {Qubit}},
	url = {https://link.aps.org/doi/10.1103/PhysRevLett.117.190503},
	doi = {10.1103/PhysRevLett.117.190503},
	language = {en},
	number = {19},
	urldate = {2025-07-08},
	journal = {Physical Review Letters},
	author = {Sank, Daniel and Chen, Zijun and Khezri, Mostafa and Kelly, J. and Barends, R. and Campbell, B. and Chen, Y. and Chiaro, B. and Dunsworth, A. and Fowler, A. and Jeffrey, E. and Lucero, E. and Megrant, A. and Mutus, J. and Neeley, M. and Neill, C. and O’Malley, P. J. J. and Quintana, C. and Roushan, P. and Vainsencher, A. and White, T. and Wenner, J. and Korotkov, Alexander N. and Martinis, John M.},
	month = nov,
	year = {2016},
	pages = {190503},
}

@article{koch_quantum_2022,
	title = {Quantum optimal control in quantum technologies. {Strategic} report on current status, visions and goals for research in {Europe}},
	volume = {9},
	issn = {2196-0763},
	url = {https://doi.org/10.1140/epjqt/s40507-022-00138-x},
	doi = {10.1140/epjqt/s40507-022-00138-x},
	number = {1},
	journal = {EPJ Quantum Technology},
	author = {Koch, Christiane P. and Boscain, Ugo and Calarco, Tommaso and Dirr, Gunther and Filipp, Stefan and Glaser, Steffen J. and Kosloff, Ronnie and Montangero, Simone and Schulte-Herbrüggen, Thomas and Sugny, Dominique and Wilhelm, Frank K.},
	month = jul,
	year = {2022},
	pages = {19},
}

@misc{mori_high-power_2025,
	title = {High-power readout of a transmon qubit using a nonlinear coupling},
	url = {http://arxiv.org/abs/2507.03642},
	doi = {10.48550/arXiv.2507.03642},
	urldate = {2025-08-14},
	publisher = {arXiv},
	author = {Mori, Cyril and Milchakov, Vladimir and D'Esposito, Francesca and Ruela, Lucas and Kumar, Shelender and Suresh, Vishnu Narayanan and Ardati, Waël and Nicolas, Dorian and Cappelli, Giulio and Ranadive, Arpit and Gal, Gwenael Le and Esposito, Martina and Ficheux, Quentin and Roch, Nicolas and Ramos, Tomás and Buisson, Olivier},
	month = aug,
	year = {2025},
	note = {arXiv:2507.03642 [quant-ph]},
}

@article{didier_fast_2015,
	title = {Fast {Quantum} {Nondemolition} {Readout} by {Parametric} {Modulation} of {Longitudinal} {Qubit}-{Oscillator} {Interaction}},
	volume = {115},
	copyright = {http://link.aps.org/licenses/aps-default-license},
	issn = {0031-9007, 1079-7114},
	url = {https://link.aps.org/doi/10.1103/PhysRevLett.115.203601},
	doi = {10.1103/PhysRevLett.115.203601},
	language = {en},
	number = {20},
	urldate = {2025-08-14},
	journal = {Physical Review Letters},
	author = {Didier, Nicolas and Bourassa, Jérôme and Blais, Alexandre},
	month = nov,
	year = {2015},
	pages = {203601},
}

@article{chatterjee_enhanced_2025,
	title = {Enhanced qubit readout via reinforcement learning},
	volume = {23},
	issn = {2331-7019},
	url = {https://link.aps.org/doi/10.1103/PhysRevApplied.23.054057},
	doi = {10.1103/PhysRevApplied.23.054057},
	language = {en},
	number = {5},
	urldate = {2025-09-10},
	journal = {Physical Review Applied},
	author = {Chatterjee, Aniket and Schwinger, Jonathan and Gao, Yvonne Y.},
	month = may,
	year = {2025},
	pages = {054057},
}

@article{wang_longitudinal_2025,
	title = {Longitudinal and {Nonlinear} {Coupling} for {High}-{Fidelity} {Readout} of a {Superconducting} {Qubit}},
	volume = {135},
	issn = {0031-9007, 1079-7114},
	url = {https://link.aps.org/doi/10.1103/98n9-13y4},
	doi = {10.1103/98n9-13y4},
	language = {en},
	number = {6},
	urldate = {2025-11-06},
	journal = {Physical Review Letters},
	author = {Wang, Can and Liu, Feng-Ming and Chen, He and Du, Yi-Fei and Ying, Chong and Wang, Jian-Wen and Huo, Yong-Heng and Peng, Cheng-Zhi and Zhu, Xiaobo and Chen, Ming-Cheng and Lu, Chao-Yang and Pan, Jian-Wei},
	month = aug,
	year = {2025},
	pages = {060803},
}

@article{krantz_quantum_2019,
	title = {A quantum engineer's guide to superconducting qubits},
	volume = {6},
	issn = {1931-9401},
	url = {https://pubs.aip.org/apr/article/6/2/021318/570326/A-quantum-engineer-s-guide-to-superconducting},
	doi = {10.1063/1.5089550},
	language = {en},
	number = {2},
	urldate = {2025-11-06},
	journal = {Applied Physics Reviews},
	author = {Krantz, P. and Kjaergaard, M. and Yan, F. and Orlando, T. P. and Gustavsson, S. and Oliver, W. D.},
	month = jun,
	year = {2019},
	pages = {021318},
}

@misc{mencia_raising_2025,
	title = {Raising the {Cavity} {Frequency} in {cQED}},
	url = {http://arxiv.org/abs/2511.22764},
	doi = {10.48550/arXiv.2511.22764},
	urldate = {2025-12-01},
	publisher = {arXiv},
	author = {Mencia, Raymond A. and Imaizumi, Taketo and Golovchanskiy, Igor A. and Lizzit, Andrea and Manucharyan, Vladimir E.},
	month = nov,
	year = {2025},
	note = {arXiv:2511.22764 [quant-ph]},
}

@article{doria_optimal_2011,
	title = {Optimal {Control} {Technique} for {Many}-{Body} {Quantum} {Dynamics}},
	volume = {106},
	copyright = {http://link.aps.org/licenses/aps-default-license},
	issn = {0031-9007, 1079-7114},
	url = {https://link.aps.org/doi/10.1103/PhysRevLett.106.190501},
	doi = {10.1103/PhysRevLett.106.190501},
	language = {en},
	number = {19},
	urldate = {2025-12-12},
	journal = {Physical Review Letters},
	author = {Doria, Patrick and Calarco, Tommaso and Montangero, Simone},
	month = may,
	year = {2011},
	pages = {190501},
}

@misc{beaulieu_fast_2026,
	title = {Fast, high-fidelity {Transmon} readout with intrinsic {Purcell} protection via nonperturbative cross-{Kerr} coupling},
	url = {http://arxiv.org/abs/2601.04975},
	doi = {10.48550/arXiv.2601.04975},
	urldate = {2026-01-09},
	publisher = {arXiv},
	author = {Beaulieu, Guillaume and Chen, Jun-Zhe and Scigliuzzo, Marco and Benhayoune-Khadraoui, Othmane and Chapple, Alex A. and Spring, Peter A. and Blais, Alexandre and Scarlino, Pasquale},
	month = jan,
	year = {2026},
	note = {arXiv:2601.04975 [quant-ph]},
}

@article{chow_simple_2011,
	title = {Simple {All}-{Microwave} {Entangling} {Gate} for {Fixed}-{Frequency} {Superconducting} {Qubits}},
	volume = {107},
	copyright = {http://link.aps.org/licenses/aps-default-license},
	issn = {0031-9007, 1079-7114},
	url = {https://link.aps.org/doi/10.1103/PhysRevLett.107.080502},
	doi = {10.1103/PhysRevLett.107.080502},
	language = {en},
	number = {8},
	urldate = {2026-01-13},
	journal = {Physical Review Letters},
	author = {Chow, Jerry M. and Córcoles, A. D. and Gambetta, Jay M. and Rigetti, Chad and Johnson, B. R. and Smolin, John A. and Rozen, J. R. and Keefe, George A. and Rothwell, Mary B. and Ketchen, Mark B. and Steffen, M.},
	month = aug,
	year = {2011},
	pages = {080502},
}

@article{tripathi_operation_2019,
	title = {Operation and intrinsic error budget of a two-qubit cross-resonance gate},
	volume = {100},
	issn = {2469-9926, 2469-9934},
	url = {https://link.aps.org/doi/10.1103/PhysRevA.100.012301},
	doi = {10.1103/PhysRevA.100.012301},
	language = {en},
	number = {1},
	urldate = {2026-01-15},
	journal = {Physical Review A},
	author = {Tripathi, Vinay and Khezri, Mostafa and Korotkov, Alexander N.},
	month = jul,
	year = {2019},
	pages = {012301},
}

@article{magesan_effective_2020,
	title = {Effective {Hamiltonian} models of the cross-resonance gate},
	volume = {101},
	issn = {2469-9926, 2469-9934},
	url = {https://link.aps.org/doi/10.1103/PhysRevA.101.052308},
	doi = {10.1103/PhysRevA.101.052308},
	language = {en},
	number = {5},
	urldate = {2026-01-15},
	journal = {Physical Review A},
	author = {Magesan, Easwar and Gambetta, Jay M.},
	month = may,
	year = {2020},
	pages = {052308},
}

@article{campaioli_quantum_2024,
	title = {Quantum {Master} {Equations}: {Tips} and {Tricks} for {Quantum} {Optics}, {Quantum} {Computing}, and {Beyond}},
	volume = {5},
	issn = {2691-3399},
	shorttitle = {Quantum {Master} {Equations}},
	url = {https://link.aps.org/doi/10.1103/PRXQuantum.5.020202},
	doi = {10.1103/PRXQuantum.5.020202},
	language = {en},
	number = {2},
	urldate = {2026-01-15},
	journal = {PRX Quantum},
	author = {Campaioli, Francesco and Cole, Jared H. and Hapuarachchi, Harini},
	month = jun,
	year = {2024},
	pages = {020202},
}

@article{sheldon_procedure_2016,
	title = {Procedure for systematically tuning up cross-talk in the cross-resonance gate},
	volume = {93},
	copyright = {http://link.aps.org/licenses/aps-default-license},
	issn = {2469-9926, 2469-9934},
	url = {https://link.aps.org/doi/10.1103/PhysRevA.93.060302},
	doi = {10.1103/PhysRevA.93.060302},
	language = {en},
	number = {6},
	urldate = {2026-01-15},
	journal = {Physical Review A},
	author = {Sheldon, Sarah and Magesan, Easwar and Chow, Jerry M. and Gambetta, Jay M.},
	month = jun,
	year = {2016},
	pages = {060302},
}

@article{rigetti_fully_2010,
	title = {Fully microwave-tunable universal gates in superconducting qubits with linear couplings and fixed transition frequencies},
	volume = {81},
	copyright = {http://link.aps.org/licenses/aps-default-license},
	issn = {1098-0121, 1550-235X},
	url = {https://link.aps.org/doi/10.1103/PhysRevB.81.134507},
	doi = {10.1103/PhysRevB.81.134507},
	language = {en},
	number = {13},
	urldate = {2026-01-20},
	journal = {Physical Review B},
	author = {Rigetti, Chad and Devoret, Michel},
	month = apr,
	year = {2010},
	pages = {134507},
}

@article{andersen_entanglement_2019,
	title = {Entanglement stabilization using ancilla-based parity detection and real-time feedback in superconducting circuits},
	volume = {5},
	copyright = {2019 The Author(s)},
	issn = {2056-6387},
	url = {https://www.nature.com/articles/s41534-019-0185-4},
	doi = {10.1038/s41534-019-0185-4},
	language = {en},
	number = {1},
	urldate = {2026-01-23},
	journal = {npj Quantum Information},
	publisher = {Nature Publishing Group},
	author = {Andersen, Christian Kraglund and Remm, Ants and Lazar, Stefania and Krinner, Sebastian and Heinsoo, Johannes and Besse, Jean-Claude and Gabureac, Mihai and Wallraff, Andreas and Eichler, Christopher},
	month = aug,
	year = {2019},
	pages = {69},
}

@article{machnes_comparing_2011,
	title = {Comparing, optimizing, and benchmarking quantum-control algorithms in a unifying programming framework},
	volume = {84},
	copyright = {http://link.aps.org/licenses/aps-default-license},
	issn = {1050-2947, 1094-1622},
	url = {https://link.aps.org/doi/10.1103/PhysRevA.84.022305},
	doi = {10.1103/PhysRevA.84.022305},
	language = {en},
	number = {2},
	urldate = {2026-01-24},
	journal = {Physical Review A},
	author = {Machnes, S. and Sander, U. and Glaser, S. J. and De Fouquières, P. and Gruslys, A. and Schirmer, S. and Schulte-Herbrüggen, T.},
	month = aug,
	year = {2011},
	pages = {022305},
}

@article{khaneja_optimal_2005,
	title = {Optimal control of coupled spin dynamics: design of {NMR} pulse sequences by gradient ascent algorithms},
	volume = {172},
	issn = {1090-7807},
	shorttitle = {Optimal control of coupled spin dynamics},
	url = {https://www.sciencedirect.com/science/article/pii/S1090780704003696},
	doi = {10.1016/j.jmr.2004.11.004},
	number = {2},
	urldate = {2026-01-24},
	journal = {Journal of Magnetic Resonance},
	author = {Khaneja, Navin and Reiss, Timo and Kehlet, Cindie and Schulte-Herbrüggen, Thomas and Glaser, Steffen J.},
	month = feb,
	year = {2005},
	pages = {296--305},
}

@article{petruhanov_grape_2023,
	title = {{GRAPE} optimization for open quantum systems with time-dependent decoherence rates driven by coherent and incoherent controls},
	volume = {56},
	issn = {1751-8113, 1751-8121},
	url = {https://iopscience.iop.org/article/10.1088/1751-8121/ace13f},
	doi = {10.1088/1751-8121/ace13f},
	number = {30},
	urldate = {2026-03-02},
	journal = {Journal of Physics A: Mathematical and Theoretical},
	author = {Petruhanov, Vadim and Pechen, Alexander},
	month = jul,
	year = {2023},
	note = {arXiv:2307.08479 [quant-ph]},
	pages = {305303},
}

@inproceedings{chen_neural_2019,
author = {Chen, Ricky T. Q. and Rubanova, Yulia and Bettencourt, Jesse and Duvenaud, David},
title = {Neural ordinary differential equations},
year = {2018},
publisher = {Curran Associates Inc.},
address = {Red Hook, NY, USA},
booktitle = {Proceedings of the 32nd International Conference on Neural Information Processing Systems},
pages = {6572–6583},
numpages = {12},
location = {Montr\'{e}al, Canada},
series = {NIPS'18}
}

@article{gautier_high-fidelity_2024,
	title = {High-fidelity control and stabilization of cat qubits},
	language = {en},
	author = {Gautier, Ronan},
    year = {2024},
}

@article{abdelhafez_gradient-based_2019,
	title = {Gradient-based optimal control of open quantum systems using quantum trajectories and automatic differentiation},
	volume = {99},
	issn = {2469-9926, 2469-9934},
	url = {https://link.aps.org/doi/10.1103/PhysRevA.99.052327},
	doi = {10.1103/PhysRevA.99.052327},
	language = {en},
	number = {5},
	urldate = {2026-03-11},
	journal = {Physical Review A},
	author = {Abdelhafez, Mohamed and Schuster, David I. and Koch, Jens},
	month = may,
	year = {2019},
	pages = {052327},
}

@misc{xie_optimal_2026,
	title = {Optimal control with flag qubits},
	url = {http://arxiv.org/abs/2603.12162},
	doi = {10.48550/arXiv.2603.12162},
	urldate = {2026-03-16},
	publisher = {arXiv},
	author = {Xie, Liang-Xu and Paula, Lui Zuccherelli de and Cai, Weizhou and Jie, Qing-Xuan and Sun, Luyan and Zou, Chang-Ling and Guo, Guang-Can and Chen, Zi-Jie and Zou, Xu-Bo},
	month = mar,
	year = {2026},
	note = {arXiv:2603.12162 [quant-ph]},
}

@misc{dixit_millimeter_2026,
	title = {Millimeter {Wave} {Readout} of a {Superconducting} {Qubit}},
	url = {http://arxiv.org/abs/2603.13837},
	doi = {10.48550/arXiv.2603.13837},
	urldate = {2026-03-17},
	publisher = {arXiv},
	author = {Dixit, Akash V. and Parrott, Zachary L. and Chunikhin, Dennis and Hauer, Bradley and Larson, Trevyn F. Q. and Teufel, John D.},
	month = mar,
	year = {2026},
	note = {arXiv:2603.13837 [quant-ph]},
}

@misc{dai_spectroscopy_2025,
	title = {Spectroscopy of drive-induced unwanted state transitions in superconducting circuits},
	url = {http://arxiv.org/abs/2506.24070},
	doi = {10.48550/arXiv.2506.24070},
	urldate = {2026-03-17},
	publisher = {arXiv},
	author = {Dai, W. and Hazra, S. and Weiss, D. K. and Kurilovich, P. D. and Connolly, T. and Babla, H. K. and Singh, S. and Joshi, V. R. and Ding, A. Z. and Parakh, P. D. and Venkatraman, J. and Xiao, X. and Frunzio, L. and Devoret, M. H.},
	month = jun,
	year = {2025},
	note = {arXiv:2506.24070 [quant-ph]
version: 1},
}

@article{schmidt_optimal_2011,
	title = {Optimal {Control} of {Open} {Quantum} {Systems}: {Cooperative} {Effects} of {Driving} and {Dissipation}},
	volume = {107},
	copyright = {http://link.aps.org/licenses/aps-default-license},
	issn = {0031-9007, 1079-7114},
	shorttitle = {Optimal {Control} of {Open} {Quantum} {Systems}},
	url = {https://link.aps.org/doi/10.1103/PhysRevLett.107.130404},
	doi = {10.1103/PhysRevLett.107.130404},
	language = {en},
	number = {13},
	urldate = {2026-03-18},
	journal = {Physical Review Letters},
	author = {Schmidt, R. and Negretti, A. and Ankerhold, J. and Calarco, T. and Stockburger, J. T.},
	month = sep,
	year = {2011},
	pages = {130404},
}

@article{goerz_charting_2017,
	title = {Charting the circuit {QED} design landscape using optimal control theory},
	volume = {3},
	copyright = {2017 The Author(s)},
	issn = {2056-6387},
	url = {https://www.nature.com/articles/s41534-017-0036-0},
	doi = {10.1038/s41534-017-0036-0},
	language = {en},
	number = {1},
	urldate = {2026-03-20},
	journal = {npj Quantum Information},
	publisher = {Nature Publishing Group},
	author = {Goerz, Michael H. and Motzoi, Felix and Whaley, K. Birgitta and Koch, Christiane P.},
	month = sep,
	year = {2017},
	pages = {37},
}

@article{nielsen_simple_2002,
	title = {A simple formula for the average gate fidelity of a quantum dynamical operation},
	volume = {303},
	issn = {0375-9601},
	url = {https://www.sciencedirect.com/science/article/pii/S0375960102012720},
	doi = {10.1016/S0375-9601(02)01272-0},
	number = {4},
	urldate = {2026-07-15},
	journal = {Physics Letters A},
	author = {Nielsen, Michael A},
	month = oct,
	year = {2002},
	pages = {249--252},
}

@article{goerz_grapejl_2025,
	title = {{GRAPE}.jl: {Gradient} {Ascent} {Pulse} {Engineering} in {Julia}},
	volume = {10},
	issn = {2475-9066},
	shorttitle = {{GRAPE}.jl},
	url = {https://joss.theoj.org/papers/10.21105/joss.08813},
	doi = {10.21105/joss.08813},
	language = {en},
	number = {115},
	urldate = {2026-03-30},
	journal = {Journal of Open Source Software},
	author = {Goerz, Michael H. and Carrasco, Sebastián C. and Marshall, Alastair and Malinovsky, Vladimir S.},
	month = nov,
	year = {2025},
	pages = {8813},
}

@article{dankert_exact_2009,
	title = {Exact and approximate unitary 2-designs and their application to fidelity estimation},
	volume = {80},
	copyright = {http://link.aps.org/licenses/aps-default-license},
	issn = {1050-2947, 1094-1622},
	url = {https://link.aps.org/doi/10.1103/PhysRevA.80.012304},
	doi = {10.1103/PhysRevA.80.012304},
	language = {en},
	number = {1},
	urldate = {2026-04-09},
	journal = {Physical Review A},
	author = {Dankert, Christoph and Cleve, Richard and Emerson, Joseph and Livine, Etera},
	month = jul,
	year = {2009},
	pages = {012304},
}

@article{matteo_short_2014,
	title = {A short introduction to unitary 2-designs},
	language = {en},
	author = {Matteo, Olivia Di},
    year = {2014},
}

@article{george_minimal_2025,
	title = {Minimal time robust control for two superconducting qubits},
	volume = {16},
	copyright = {2025 The Author(s)},
	issn = {2045-2322},
	url = {https://www.nature.com/articles/s41598-025-32747-8},
	doi = {10.1038/s41598-025-32747-8},
	language = {en},
	number = {1},
	urldate = {2026-05-26},
	journal = {Scientific Reports},
	publisher = {Nature Publishing Group},
	author = {George, Niril and Allen, Joseph L. and Kosut, Robert and Ginossar, Eran},
	month = dec,
	year = {2025},
	pages = {2773},
}

@article{kosut_robust_2013,
	title = {Robust control of quantum gates via sequential convex programming},
	volume = {88},
	copyright = {http://link.aps.org/licenses/aps-default-license},
	issn = {1050-2947, 1094-1622},
	url = {https://link.aps.org/doi/10.1103/PhysRevA.88.052326},
	doi = {10.1103/PhysRevA.88.052326},
	language = {en},
	number = {5},
	urldate = {2026-05-26},
	journal = {Physical Review A},
	author = {Kosut, Robert L. and Grace, Matthew D. and Brif, Constantin},
	month = nov,
	year = {2013},
	pages = {052326},
}

@phdthesis{remm_implementing_2023,
	title = {Implementing {Surface} {Codes} with {Superconducting} {Circuits}},
	url = {http://hdl.handle.net/20.500.11850/652933},
	doi = {10.3929/ETHZ-B-000652933},
	language = {en},
	urldate = {2026-05-26},
	school = {ETH Zurich},
	author = {Remm, Ants},
	collaborator = {{Wallraff, Andreas} and {Grimm, Alexander}},
	year = {2023},
	note = {Artwork Size: 188 p.
Medium: application/pdf
Pages: 188 p.},
}

@article{goutte_low-rank_2026,
	title = {Low-rank optimal control of quantum devices},
	volume = {8},
	issn = {2643-1564},
	url = {https://link.aps.org/doi/10.1103/v383-2vss},
	doi = {10.1103/v383-2vss},
	language = {en},
	number = {1},
	urldate = {2026-06-07},
	journal = {Physical Review Research},
	author = {Goutte, Leo and Savona, Vincenzo},
	month = jan,
	year = {2026},
	pages = {013085},
}

@article{leung_speedup_2017,
	title = {Speedup for quantum optimal control from automatic differentiation based on graphics processing units},
	volume = {95},
	copyright = {http://link.aps.org/licenses/aps-default-license},
	issn = {2469-9926, 2469-9934},
	url = {http://link.aps.org/doi/10.1103/PhysRevA.95.042318},
	doi = {10.1103/PhysRevA.95.042318},
	language = {en},
	number = {4},
	urldate = {2026-06-16},
	journal = {Physical Review A},
	author = {Leung, Nelson and Abdelhafez, Mohamed and Koch, Jens and Schuster, David},
	month = apr,
	year = {2017},
	pages = {042318},
}

@article{lu_optimal_2024,
	title = {Optimal control of large quantum systems: assessing memory and runtime performance of {GRAPE}},
	volume = {8},
	issn = {2399-6528},
	shorttitle = {Optimal control of large quantum systems},
	url = {https://doi.org/10.1088/2399-6528/ad22e5},
	doi = {10.1088/2399-6528/ad22e5},
	language = {en},
	number = {2},
	urldate = {2026-06-16},
	journal = {Journal of Physics Communications},
	publisher = {IOP Publishing},
	author = {Lu, Yunwei and Joshi, Sandeep and San Dinh, Vinh and Koch, Jens},
	month = feb,
	year = {2024},
	pages = {025002},
}

@article{goerz_quantum_2022,
	title = {Quantum {Optimal} {Control} via {Semi}-{Automatic} {Differentiation}},
	volume = {6},
	url = {https://quantum-journal.org/papers/q-2022-12-07-871/},
	doi = {10.22331/q-2022-12-07-871},
	language = {en-GB},
	urldate = {2026-07-15},
	journal = {Quantum},
	publisher = {Verein zur Förderung des Open Access Publizierens in den Quantenwissenschaften},
	author = {Goerz, Michael H. and Carrasco, Sebastián C. and Malinovsky, Vladimir S.},
	month = dec,
	year = {2022},
	pages = {871},
}

@article{boutin_resonator_2017,
	title = {Resonator reset in circuit {QED} by optimal control for large open quantum systems},
	volume = {96},
	copyright = {https://link.aps.org/licenses/aps-default-license},
	issn = {2469-9926, 2469-9934},
	url = {https://link.aps.org/doi/10.1103/PhysRevA.96.042315},
	doi = {10.1103/PhysRevA.96.042315},
	language = {en},
	number = {4},
	urldate = {2026-06-16},
	journal = {Physical Review A},
	author = {Boutin, Samuel and Andersen, Christian Kraglund and Venkatraman, Jayameenakshi and Ferris, Andrew J. and Blais, Alexandre},
	month = oct,
	year = {2017},
	pages = {042315},
}

@article{boscain_introduction_2021,
	title = {Introduction to the {Pontryagin} {Maximum} {Principle} for {Quantum} {Optimal} {Control}},
	volume = {2},
	issn = {2691-3399},
	url = {https://link.aps.org/doi/10.1103/PRXQuantum.2.030203},
	doi = {10.1103/PRXQuantum.2.030203},
	language = {en},
	number = {3},
	urldate = {2026-06-19},
	journal = {PRX Quantum},
	author = {Boscain, U. and Sigalotti, M. and Sugny, D.},
	month = sep,
	year = {2021},
	pages = {030203},
}

@article{geher_reset_2025,
	title = {To reset, or not to reset—that is the question},
	volume = {11},
	copyright = {2025 The Author(s)},
	issn = {2056-6387},
	url = {https://www.nature.com/articles/s41534-025-00998-y},
	doi = {10.1038/s41534-025-00998-y},
	language = {en},
	number = {1},
	urldate = {2026-06-26},
	journal = {npj Quantum Information},
	publisher = {Nature Publishing Group},
	author = {Gehér, György P. and Jastrzebski, Marcin and Campbell, Earl T. and Crawford, Ophelia},
	month = mar,
	year = {2025},
	pages = {39},
}

@misc{ali_surface-code_2025,
	title = {Surface-code {Superconducting} {Quantum} {Processors}: {From} {Calibration} {To} {Logical} {Performance}},
	shorttitle = {Surface-code {Superconducting} {Quantum} {Processors}},
	url = {http://arxiv.org/abs/2504.17082},
	doi = {10.48550/arXiv.2504.17082},
	urldate = {2026-06-28},
	publisher = {arXiv},
	author = {Ali, Hany},
	month = apr,
	year = {2025},
	note = {arXiv:2504.17082 [quant-ph]},
}

@article{marques_logical-qubit_2022,
	title = {Logical-qubit operations in an error-detecting surface code},
	volume = {18},
	copyright = {2021 The Author(s), under exclusive licence to Springer Nature Limited},
	issn = {1745-2481},
	url = {https://www.nature.com/articles/s41567-021-01423-9},
	doi = {10.1038/s41567-021-01423-9},
	language = {en},
	number = {1},
	urldate = {2026-06-28},
	journal = {Nature Physics},
	publisher = {Nature Publishing Group},
	author = {Marques, J. F. and Varbanov, B. M. and Moreira, M. S. and Ali, H. and Muthusubramanian, N. and Zachariadis, C. and Battistel, F. and Beekman, M. and Haider, N. and Vlothuizen, W. and Bruno, A. and Terhal, B. M. and DiCarlo, L.},
	month = jan,
	year = {2022},
	pages = {80--86},
}

@article{kjaergaard_superconducting_2020,
	title = {Superconducting {Qubits}: {Current} {State} of {Play}},
	volume = {11},
	issn = {1947-5454, 1947-5462},
	shorttitle = {Superconducting {Qubits}},
	url = {https://www.annualreviews.org/content/journals/10.1146/annurev-conmatphys-031119-050605},
	doi = {10.1146/annurev-conmatphys-031119-050605},
	language = {en},
	number = {Volume 11, 2020},
	urldate = {2026-07-15},
	journal = {Annual Review of Condensed Matter Physics},
	publisher = {Annual Reviews},
	author = {Kjaergaard, Morten and Schwartz, Mollie E. and Braumüller, Jochen and Krantz, Philip and Wang, Joel I.-J. and Gustavsson, Simon and Oliver, William D.},
	month = mar,
	year = {2020},
	pages = {369--395},
}

@article{chow_implementing_2014,
	title = {Implementing a strand of a scalable fault-tolerant quantum computing fabric},
	volume = {5},
	copyright = {2014 Springer Nature Limited},
	issn = {2041-1723},
	url = {https://www.nature.com/articles/ncomms5015},
	doi = {10.1038/ncomms5015},
	language = {en},
	number = {1},
	urldate = {2026-07-15},
	journal = {Nature Communications},
	publisher = {Nature Publishing Group},
	author = {Chow, Jerry M. and Gambetta, Jay M. and Magesan, Easwar and Abraham, David W. and Cross, Andrew W. and Johnson, B. R. and Masluk, Nicholas A. and Ryan, Colm A. and Smolin, John A. and Srinivasan, Srikanth J. and Steffen, M.},
	month = jun,
	year = {2014},
	pages = {4015},
}

@article{saira_entanglement_2014,
	title = {Entanglement {Genesis} by {Ancilla}-{Based} {Parity} {Measurement} in {2D} {Circuit} {QED}},
	volume = {112},
	copyright = {http://link.aps.org/licenses/aps-default-license},
	issn = {0031-9007, 1079-7114},
	url = {https://link.aps.org/doi/10.1103/PhysRevLett.112.070502},
	doi = {10.1103/PhysRevLett.112.070502},
	language = {en},
	number = {7},
	urldate = {2026-06-29},
	journal = {Physical Review Letters},
	author = {Saira, O.-P. and Groen, J. P. and Cramer, J. and Meretska, M. and De Lange, G. and DiCarlo, L.},
	month = feb,
	year = {2014},
	pages = {070502},
}

@article{chen_robust_2025,
	title = {Robust and optimal control of open quantum systems},
	volume = {11},
	url = {https://www.science.org/doi/10.1126/sciadv.adr0875},
	doi = {10.1126/sciadv.adr0875},
	number = {9},
	urldate = {2026-07-02},
	journal = {Science Advances},
	publisher = {American Association for the Advancement of Science},
	author = {Chen, Zi-Jie and Huang, Hongwei and Sun, Lida and Jie, Qing-Xuan and Zhou, Jie and Hua, Ziyue and Xu, Yifang and Wang, Weiting and Guo, Guang-Can and Zou, Chang-Ling and Sun, Luyan and Zou, Xu-Bo},
	month = feb,
	year = {2025},
	pages = {eadr0875},
}

@article{gravina_adaptive_2024,
	title = {Adaptive variational low-rank dynamics for open quantum systems},
	volume = {6},
	issn = {2643-1564},
	url = {https://link.aps.org/doi/10.1103/PhysRevResearch.6.023072},
	doi = {10.1103/PhysRevResearch.6.023072},
	language = {en},
	number = {2},
	urldate = {2025-01-24},
	journal = {Physical Review Research},
	author = {Gravina, Luca and Savona, Vincenzo},
	month = apr,
	year = {2024},
	pages = {023072},
}

@article{joubert-doriol_non-stochastic_2014,
	title = {Non-stochastic matrix {Schrödinger} equation for open systems},
	volume = {141},
	issn = {0021-9606, 1089-7690},
	url = {https://pubs.aip.org/jcp/article/141/23/234112/194230/Non-stochastic-matrix-Schrodinger-equation-for},
	doi = {10.1063/1.4903829},
	language = {en},
	number = {23},
	urldate = {2025-05-06},
	journal = {The Journal of Chemical Physics},
	author = {Joubert-Doriol, Loïc and Ryabinkin, Ilya G. and Izmaylov, Artur F.},
	month = dec,
	year = {2014},
	pages = {234112},
}

@article{santos_low-rank_2025,
	title = {Low-{Rank} {Variational} {Quantum} {Algorithm} for the {Dynamics} of {Open} {Quantum} {Systems}},
	volume = {9},
	url = {https://quantum-journal.org/papers/q-2025-02-04-1620/},
	doi = {10.22331/q-2025-02-04-1620},
	language = {en-GB},
	urldate = {2026-07-15},
	journal = {Quantum},
	publisher = {Verein zur Förderung des Open Access Publizierens in den Quantenwissenschaften},
	author = {Santos, Sara and Song, Xinyu and Savona, Vincenzo},
	month = feb,
	year = {2025},
	pages = {1620},
}

@article{joubert-doriol_problem-free_2015,
	title = {Problem-free time-dependent variational principle for open quantum systems},
	volume = {142},
	issn = {0021-9606, 1089-7690},
	url = {https://pubs.aip.org/jcp/article/142/13/134107/901966/Problem-free-time-dependent-variational-principle},
	doi = {10.1063/1.4916384},
	language = {en},
	number = {13},
	urldate = {2025-05-07},
	journal = {The Journal of Chemical Physics},
	author = {Joubert-Doriol, Loïc and Izmaylov, Artur F.},
	month = apr,
	year = {2015},
	pages = {134107},
}

@article{le_bris_low-rank_2013,
	title = {Low-rank numerical approximations for high-dimensional {Lindblad} equations},
	volume = {87},
	url = {https://link.aps.org/doi/10.1103/PhysRevA.87.022125},
	doi = {10.1103/PhysRevA.87.022125},
	number = {2},
	urldate = {2025-07-07},
	journal = {Physical Review A},
	publisher = {American Physical Society},
	author = {Le Bris, C. and Rouchon, P.},
	month = feb,
	year = {2013},
	pages = {022125},
}

@article{le_bris_adaptive_2015,
	title = {Adaptive low-rank approximation and denoised {Monte} {Carlo} approach for high-dimensional {Lindblad} equations},
	volume = {92},
	copyright = {http://link.aps.org/licenses/aps-default-license},
	issn = {1050-2947, 1094-1622},
	url = {https://link.aps.org/doi/10.1103/PhysRevA.92.062126},
	doi = {10.1103/PhysRevA.92.062126},
	language = {en},
	number = {6},
	urldate = {2025-11-04},
	journal = {Physical Review A},
	author = {Le Bris, C. and Rouchon, P. and Roussel, J.},
	month = dec,
	year = {2015},
	pages = {062126},
}

@article{mccaul_fast_2021,
	title = {Fast computation of dissipative quantum systems with ensemble rank truncation},
	volume = {3},
	issn = {2643-1564},
	url = {https://link.aps.org/doi/10.1103/PhysRevResearch.3.013017},
	doi = {10.1103/PhysRevResearch.3.013017},
	language = {en},
	number = {1},
	urldate = {2026-06-07},
	journal = {Physical Review Research},
	author = {McCaul, Gerard and Jacobs, Kurt and Bondar, Denys I.},
	month = jan,
	year = {2021},
	pages = {013017},
}

@book{breuer_theory_2002,
	address = {Oxford ; New York},
	title = {The theory of open quantum systems},
	isbn = {978-0-19-852063-4},
	language = {en},
	publisher = {Oxford University Press},
	author = {Breuer, Heinz-Peter and Petruccione, F.},
	year = {2002},
	note = {OCLC: ocm49872077},
}

@article{ferrari_dissipative_2025,
	title = {Dissipative quantum chaos unveiled by stochastic quantum trajectories},
	volume = {7},
	issn = {2643-1564},
	url = {https://link.aps.org/doi/10.1103/PhysRevResearch.7.013276},
	doi = {10.1103/PhysRevResearch.7.013276},
	language = {en},
	number = {1},
	urldate = {2025-07-08},
	journal = {Physical Review Research},
	author = {Ferrari, Filippo and Gravina, Luca and Eeltink, Debbie and Scarlino, Pasquale and Savona, Vincenzo and Minganti, Fabrizio},
	month = mar,
	year = {2025},
	pages = {013276},
}

@article{weimer_simulation_2021,
	title = {Simulation methods for open quantum many-body systems},
	volume = {93},
	issn = {0034-6861, 1539-0756},
	url = {https://link.aps.org/doi/10.1103/RevModPhys.93.015008},
	doi = {10.1103/RevModPhys.93.015008},
	language = {en},
	number = {1},
	urldate = {2025-07-08},
	journal = {Reviews of Modern Physics},
	author = {Weimer, Hendrik and Kshetrimayum, Augustine and Orús, Román},
	month = mar,
	year = {2021},
	pages = {015008},
}

@article{dalibard_wave_function_1992,
  title = {Wave-function approach to dissipative processes in quantum optics},
  author = {Dalibard, Jean and Castin, Yvan and M\o{}lmer, Klaus},
  journal = {Phys. Rev. Lett.},
  volume = {68},
  issue = {5},
  pages = {580--583},
  numpages = {0},
  year = {1992},
  month = {Feb},
  publisher = {American Physical Society},
  doi = {10.1103/PhysRevLett.68.580},
  url = {https://link.aps.org/doi/10.1103/PhysRevLett.68.580}
}

@article{molmer_monte_1996,
	title = {Monte {Carlo} wavefunctions in quantum optics},
	volume = {8},
	issn = {1355-5111},
	url = {https://doi.org/10.1088/1355-5111/8/1/007},
	doi = {10.1088/1355-5111/8/1/007},
	language = {en},
	number = {1},
	urldate = {2026-07-03},
	journal = {Quantum and Semiclassical Optics: Journal of the European Optical Society Part B},
	author = {Mølmer, Klaus and Castin, Yvan},
	month = feb,
	year = {1996},
	pages = {49},
}

@article{guo_locally_2024,
	title = {Locally purified density operators for noisy quantum circuits},
	volume = {41},
	issn = {0256-307X, 1741-3540},
	url = {https://iopscience.iop.org/article/10.1088/0256-307X/41/12/120302},
	doi = {10.1088/0256-307X/41/12/120302},
	language = {en},
	number = {12},
	urldate = {2025-10-31},
	journal = {Chinese Physics Letters},
	author = {Guo, Yuchen and Yang, Shuo},
	month = dec,
	year = {2024},
	pages = {120302},
}

@article{werner_positive_2016,
	title = {Positive {Tensor} {Network} {Approach} for {Simulating} {Open} {Quantum} {Many}-{Body} {Systems}},
	volume = {116},
	copyright = {http://link.aps.org/licenses/aps-default-license},
	issn = {0031-9007, 1079-7114},
	url = {https://link.aps.org/doi/10.1103/PhysRevLett.116.237201},
	doi = {10.1103/PhysRevLett.116.237201},
	language = {en},
	number = {23},
	urldate = {2025-11-03},
	journal = {Physical Review Letters},
	author = {Werner, A. H. and Jaschke, D. and Silvi, P. and Kliesch, M. and Calarco, T. and Eisert, J. and Montangero, S.},
	month = jun,
	year = {2016},
	pages = {237201},
}

@article{verstraete_matrix_2004,
	title = {Matrix {Product} {Density} {Operators}: {Simulation} of {Finite}-{Temperature} and {Dissipative} {Systems}},
	volume = {93},
	copyright = {http://link.aps.org/licenses/aps-default-license},
	issn = {0031-9007, 1079-7114},
	shorttitle = {Matrix {Product} {Density} {Operators}},
	url = {https://link.aps.org/doi/10.1103/PhysRevLett.93.207204},
	doi = {10.1103/PhysRevLett.93.207204},
	language = {en},
	number = {20},
	urldate = {2025-11-24},
	journal = {Physical Review Letters},
	author = {Verstraete, F. and García-Ripoll, J. J. and Cirac, J. I.},
	month = nov,
	year = {2004},
	pages = {207204},
}

@inproceedings{kingma_adam_2015,
  author    = {Kingma, Diederik P. and Ba, Jimmy},
  title     = {Adam: A Method for Stochastic Optimization},
  booktitle = {International Conference on Learning Representations (ICLR)},
  year      = {2015},
  url       = {https://arxiv.org/abs/1412.6980}
}

@inproceedings{moses_instead_2020,
 author = {Moses, William and Churavy, Valentin},
 booktitle = {Advances in Neural Information Processing Systems},
 editor = {H. Larochelle and M. Ranzato and R. Hadsell and M. F. Balcan and H. Lin},
 pages = {12472--12485},
 publisher = {Curran Associates, Inc.},
 title = {Instead of Rewriting Foreign Code for Machine Learning, Automatically Synthesize Fast Gradients},
 url = {https://proceedings.neurips.cc/paper/2020/file/9332c513ef44b682e9347822c2e457ac-Paper.pdf},
 volume = {33},
 year = {2020}
}

@inproceedings{moses_reverse-mode_2021,
 author = {Moses, William S. and Churavy, Valentin and Paehler, Ludger and H\"{u}ckelheim, Jan and Narayanan, Sri Hari Krishna and Schanen, Michel and Doerfert, Johannes},
 title = {Reverse-Mode Automatic Differentiation and Optimization of GPU Kernels via Enzyme},
 year = {2021},
 isbn = {9781450384421},
 publisher = {Association for Computing Machinery},
 address = {New York, NY, USA},
 url = {https://doi.org/10.1145/3458817.3476165},
 doi = {10.1145/3458817.3476165},
 booktitle = {Proceedings of the International Conference for High Performance Computing, Networking, Storage and Analysis},
 articleno = {61},
 numpages = {16},
 location = {St. Louis, Missouri},
 series = {SC '21}
}

@article{mercurio_quantum_2025,
  doi = {10.22331/q-2025-09-29-1866},
  url = {https://doi.org/10.22331/q-2025-09-29-1866},
  title = {Quantum{T}oolbox.jl: {A}n efficient {J}ulia framework for simulating open quantum systems},
  author = {Mercurio, Alberto and Huang, Yi-Te and Cai, Li-Xun and Chen, Yueh-Nan and Savona, Vincenzo and Nori, Franco},
  journal = {{Quantum}},
  issn = {2521-327X},
  publisher = {{Verein zur F{\"{o}}rderung des Open Access Publizierens in den Quantenwissenschaften}},
  volume = {9},
  pages = {1866},
  month = sep,
  year = {2025}
}

@article{klimov_discrete_2009,
	title = {Discrete phase-space structure of \textit{n}-qubit mutually unbiased bases},
	volume = {324},
	issn = {0003-4916},
	url = {https://www.sciencedirect.com/science/article/pii/S0003491608001541},
	doi = {10.1016/j.aop.2008.10.003},
	number = {1},
	urldate = {2026-07-08},
	journal = {Annals of Physics},
	author = {Klimov, A. B. and Romero, J. L. and Björk, G. and Sánchez-Soto, L. L.},
	month = jan,
	year = {2009},
	pages = {53--72},
}

@article{klimov_geometrical_2007,
	title = {Geometrical approach to mutually unbiased bases},
	volume = {40},
	issn = {1751-8121},
	url = {https://doi.org/10.1088/1751-8113/40/14/014},
	doi = {10.1088/1751-8113/40/14/014},
	language = {en},
	number = {14},
	urldate = {2026-07-15},
	journal = {Journal of Physics A: Mathematical and Theoretical},
	author = {Klimov, Andrei B and Romero, José L and Björk, Gunnar and Sánchez-Soto, Luis L},
	month = mar,
	year = {2007},
	pages = {3987},
}

@misc{goutte_zenodo_2026,
  author       = {Goutte, Leo},
  title        = {LROC.jl: Low-rank optimal control},
  month        = jul,
  year         = 2026,
  publisher    = {Zenodo},
  version      = {v0.1.0-alpha},
  doi          = {10.5281/zenodo.21372277},
  url          = {https://doi.org/10.5281/zenodo.21372277},
}

@article{genois_quantum_2025,
  title = {Quantum optimal control of superconducting qubits based on machine-learning characterization},
  author = {Genois, \'Elie and Stevenson, Noah J. and Goss, Noah and Siddiqi, Irfan and Blais, Alexandre},
  journal = {Phys. Rev. Appl.},
  volume = {24},
  issue = {3},
  pages = {034073},
  numpages = {16},
  year = {2025},
  month = {Sep},
  publisher = {American Physical Society},
  doi = {10.1103/d9yg-d3qr},
  url = {https://link.aps.org/doi/10.1103/d9yg-d3qr}
}

@article{rackauckas_differential_2017,
  title={Differentialequations.jl--a performant and feature-rich ecosystem for solving differential equations in julia},
  author={Rackauckas, Christopher and Nie, Qing},
  journal={Journal of Open Research Software},
  volume={5},
  number={1},
  pages={15},
  year={2017},
  publisher={Ubiquity Press}
}

@misc{koor_ashort_2023,
      title={A short tutorial on Wirtinger Calculus with applications in quantum information}, 
      author={Kelvin Koor and Yixian Qiu and Leong Chuan Kwek and Patrick Rebentrost},
      year={2023},
      eprint={2312.04858},
      archivePrefix={arXiv},
      primaryClass={quant-ph},
      url={https://arxiv.org/abs/2312.04858}, 
}

@article{baum_experimental_2021,
  title = {Experimental Deep Reinforcement Learning for Error-Robust Gate-Set Design on a Superconducting Quantum Computer},
  author = {Baum, Yuval and Amico, Mirko and Howell, Sean and Hush, Michael and Liuzzi, Maggie and Mundada, Pranav and Merkh, Thomas and Carvalho, Andre R.R. and Biercuk, Michael J.},
  journal = {PRX Quantum},
  volume = {2},
  issue = {4},
  pages = {040324},
  numpages = {12},
  year = {2021},
  month = {Nov},
  publisher = {American Physical Society},
  doi = {10.1103/PRXQuantum.2.040324},
  url = {https://link.aps.org/doi/10.1103/PRXQuantum.2.040324}
}

@article{caneva_chopped_2011,
  title = {Chopped random-basis quantum optimization},
  author = {Caneva, Tommaso and Calarco, Tommaso and Montangero, Simone},
  journal = {Phys. Rev. A},
  volume = {84},
  issue = {2},
  pages = {022326},
  numpages = {9},
  year = {2011},
  month = {Aug},
  publisher = {American Physical Society},
  doi = {10.1103/PhysRevA.84.022326},
  url = {https://link.aps.org/doi/10.1103/PhysRevA.84.022326}
}

@article{tsitouras_runge-kutta_2011,
	title = {Runge-{Kutta} pairs of order 5(4) satisfying only the first column simplifying assumption},
	volume = {62},
	url = {https://dl.acm.org/doi/10.1016/j.camwa.2011.06.002},
	doi = {10.1016/j.camwa.2011.06.002},
	number = {2},
	urldate = {2026-07-15},
	journal = {Computers \& Mathematics with Applications},
	publisher = {Pergamon Press, Inc.},
	author = {Tsitouras, Ch. and {View Profile}},
	month = jul,
	year = {2011},
	pages = {770--775},
}

@article{zhu_cross-platform_2022,
	title = {Cross-platform comparison of arbitrary quantum states},
	volume = {13},
	copyright = {2022 The Author(s)},
	issn = {2041-1723},
	url = {https://www.nature.com/articles/s41467-022-34279-5},
	doi = {10.1038/s41467-022-34279-5},
	language = {en},
	number = {1},
	urldate = {2026-07-15},
	journal = {Nature Communications},
	publisher = {Nature Publishing Group},
	author = {Zhu, D. and Cian, Z. P. and Noel, C. and Risinger, A. and Biswas, D. and Egan, L. and Zhu, Y. and Green, A. M. and Alderete, C. Huerta and Nguyen, N. H. and Wang, Q. and Maksymov, A. and Nam, Y. and Cetina, M. and Linke, N. M. and Hafezi, M. and Monroe, C.},
	month = nov,
	year = {2022},
	pages = {6620},
}

@article{arute_quantum_2019,
	title = {Quantum supremacy using a programmable superconducting processor},
	volume = {574},
	copyright = {2019 The Author(s), under exclusive licence to Springer Nature Limited},
	issn = {1476-4687},
	url = {https://www.nature.com/articles/s41586-019-1666-5},
	doi = {10.1038/s41586-019-1666-5},
	language = {en},
	number = {7779},
	urldate = {2026-07-15},
	journal = {Nature},
	publisher = {Nature Publishing Group},
	author = {Arute, Frank and Arya, Kunal and Babbush, Ryan and Bacon, Dave and Bardin, Joseph C. and Barends, Rami and Biswas, Rupak and Boixo, Sergio and Brandao, Fernando G. S. L. and Buell, David A. and Burkett, Brian and Chen, Yu and Chen, Zijun and Chiaro, Ben and Collins, Roberto and Courtney, William and Dunsworth, Andrew and Farhi, Edward and Foxen, Brooks and Fowler, Austin and Gidney, Craig and Giustina, Marissa and Graff, Rob and Guerin, Keith and Habegger, Steve and Harrigan, Matthew P. and Hartmann, Michael J. and Ho, Alan and Hoffmann, Markus and Huang, Trent and Humble, Travis S. and Isakov, Sergei V. and Jeffrey, Evan and Jiang, Zhang and Kafri, Dvir and Kechedzhi, Kostyantyn and Kelly, Julian and Klimov, Paul V. and Knysh, Sergey and Korotkov, Alexander and Kostritsa, Fedor and Landhuis, David and Lindmark, Mike and Lucero, Erik and Lyakh, Dmitry and Mandrà, Salvatore and McClean, Jarrod R. and McEwen, Matthew and Megrant, Anthony and Mi, Xiao and Michielsen, Kristel and Mohseni, Masoud and Mutus, Josh and Naaman, Ofer and Neeley, Matthew and Neill, Charles and Niu, Murphy Yuezhen and Ostby, Eric and Petukhov, Andre and Platt, John C. and Quintana, Chris and Rieffel, Eleanor G. and Roushan, Pedram and Rubin, Nicholas C. and Sank, Daniel and Satzinger, Kevin J. and Smelyanskiy, Vadim and Sung, Kevin J. and Trevithick, Matthew D. and Vainsencher, Amit and Villalonga, Benjamin and White, Theodore and Yao, Z. Jamie and Yeh, Ping and Zalcman, Adam and Neven, Hartmut and Martinis, John M.},
	month = oct,
	year = {2019},
	pages = {505--510},
}

\appendix

\section{\label{app:formalism}Details and derivation of LROC algorithm}

In this appendix we collect the derivation details underlying the main-text results of Sec.~\ref{sec:formalism}. The derivation follows the application of the Weak Pontryagin maximum principle~\cite{boscain_introduction_2021} to the low-rank evolution $\dot{\m} = \mathbf f(\m, \mathbf u)$, where $\mathbf u$ is the vector of parameters $\ukj$. A convenient starting point is the functional
\begin{align}\label{eq:lagrangian}
\begin{split}
    \Lambda[\w, \mathbf u] =& \Phi\left[\m(T) \right] + \int_0^T \phi[\m(t), \mathbf u(t)]\, \diff{t} \\
    & - 2 \mathrm{Re}\int_0^T \langle \bm\lambda(t), \dot{\m}(t) - \mathbf f[\m(t), \mathbf u(t)]\rangle \ \diff{t},
\end{split}
\end{align}
the promotion of Eq.~\eqref{eq:loss} which formally imposes Eq.~\eqref{eq:nosse} as a dynamical constraint (the factor of $2$ and the real part originate from imposing the same on the evolution of $\m^\dagger$). 
Here, $\w$ is a Lagrange multiplier defined on $[0, T]$ known as the adjoint state, or \emph{sensitivity}. The inner product is defined as $\langle A, B \rangle = \mathrm{tr}(A^\dagger B)$.

The inclusion of $\w$ in Eq.~\eqref{eq:lagrangian} allows for a simple derivative with respect to the control parameters, whose sole dependence now lies in the term $\mathbf f(\m, \mathbf u)$. For piecewise-constant controls, the gradient with respect to $\ukj$ may be written as
\begin{align}\label{eq:gradient_integral}
\begin{split}
        \frac{\partial \Lambda}{\partial \ukj} &= 2 \Re\int_0^T \left\langle{\w, \frac{\partial \mathbf f}{\partial \ukj}} \right\rangle dt \\
    &= 2\Im \int_{t_j}^{t_{j+1}} \left\langle{\wj, H_k \mj}\right\rangle dt \\
    &\simeq 2\Delta t\, \Im\, \left\langle{\wj H_k \mj}\right\rangle,
\end{split}
\end{align}
where the last step incurs an $\mathcal O(\Delta t^2)$ error. In practice, one may instead compute the gradient using the trapezoidal rule for which the error scales as $\mathcal O (\Delta t^3)$, or integrate the gradient together with the adjoint state by employing an augmented state method as explained later in Appendix~\ref{app:implementation}.

We consider the variation $\delta \Lambda$ in $\Lambda$ to first order to be due to the variation $\delta \m$ in $\m$ and its Hermitian conjugate $\m^\dagger$. Note that $\w$ does not change. The adjoint equation of motion and terminal condition is found by imposing $\delta \Lambda = 0$, integrating $-\langle \w, \delta \dot{\m}\rangle$ by parts, and collecting bulk and boundary terms. The adjoint state is found to satisfy
\begin{equation}\label{eq:adjoint_eom_app}
    \dot{\w} = -\left(\frac{\partial \mathbf f}{\partial \m}\right)^\dagger \w - \frac{\partial \phi}{\partial \m^*},
\end{equation}
with terminal condition 
\begin{equation}
    \w(T) = \frac{\partial \Phi }{\partial \m^*}.
\end{equation}
Linearizing the forward equation of motion Eq.~\eqref{eq:nosse} yields
\begin{align}\label{eq:dfdm}
\left(\frac{\partial \mathbf f}{\partial \m}\right)^\dagger \w = \I H \w + J_{\mathcal{O}}^\dagger \w + J_{\mu}^\dagger \w.
\end{align}
The first term is the usual backward-in-time dynamics while the remaining terms are the contribution due to dissipation and norm-preservation defined, respectively, as
\begin{widetext}
    \begin{subequations}
        \begin{gather}
        J_{\mathcal{O}}^\dagger \w = \frac{1}{2} \sum_s \left[ L_s^\dagger \w A_s  - L_s^\dagger L_s \w + \left(L_s \m- \m A_s\right) B_s^\dagger + L_s^\dagger \m B_s - \m B_s A_s^\dagger \right] ,\\
        J_{\mu}^\dagger \w = \mu \w + \nu \frac{J_{\mathcal{O}}^\dagger \m + \mathcal{O}[\m] - 2\mu \m}{\mathrm{tr}(\m^\dagger \m)}.
        \end{gather}
    \end{subequations}
\end{widetext}
Moreover, we define $A_s =  \m^+ L_s \m$ and $B_s = (\m^\dagger \m)^{-1} \bm\lambda^\dagger L_s \m$, and $\nu = \mathrm{Re}\,\mathrm{tr}(\w^\dagger \m)$. Note that Eq.~\eqref{eq:dfdm} is linear in $\w$, and its coefficients are determined by the forward trajectories $\m$. 

The terms in Eq.~\eqref{eq:dfdm} admit a transparent geometric interpretation. Let $\Pi = \m\m^+$ be the orthogonal projector onto the column space of $\m$, i.e.\ the instantaneous support of $\rho$ in the low-rank manifold. Then $A_s = \m^+ L_s\m$ represents the jump operator $L_s$ within this support, with $\m A_s = \Pi\, L_s\m$ the on-manifold part of the jumped state. The term $B_s = (\m^\dagger\m)^{-1}\bm\lambda^\dagger L_s\m$ is its adjoint counterpart, obtained by replacing $\m$ with $\bm\lambda$ in the bra, and arises purely from differentiating the pseudo-inverse.

The terms without $B_s$ are the transpose of the forward generator at fixed support: $iH\bm\lambda$ propagates the sensitivity under the reversed Hamiltonian flow, $ L_s^\dagger\bm\lambda A_s$ is the reversed quantum jump term, and $- L_s^\dagger L_s\bm\lambda$ is the same non-Hermitian decay as in the forward map. Under the backward integration it carries an overall negative sign, contributing to the stability of the adjoint pass. The $B_s$ terms are the closed-form derivative of the manifold projection: of these, $(L_s\m - \m A_s)B_s^\dagger = ( 1 - \Pi)L_s\m\,B_s^\dagger$ is the leakage of the jump off the rank-$M$ subspace and vanishes when the approximation is exact, while $L_s^\dagger\m B_s - \m B_s A_s^\dagger$ complete the derivative. The remainder of the terms in $J_\mu^\dagger \w$ enforce trace-preservation of the forward state in the backward pass. Together, they are precisely what automatic differentiation through Eq.~\eqref{eq:nosse} reconstructs.

\section{Select loss function contributions}\label{app:losses}

A practical strength of the LROC formalism is its modularity. A given optimization objective enters purely into the adjoint dynamics and only through two quantities: the terminal derivative $\partial\Phi/\partial\m^*$ which sets the boundary condition $\bm\lambda(T)$, and the running derivative $\partial\phi/\partial\m^*$ which acts as a source in Eq.~\eqref{eq:adjoint_eom}. Adapting LROC to a new cost therefore requires only these two derivatives, obtained by differentiating the functional with respect to $\m^*$ while treating $\m$ and $\m^*$ as independent~\cite{koor_ashort_2023}.

In this section, we present the most common loss function contributions and their respective gradients to facilitate the use of the LROC procedure. The elementary building block is the (unnormalized) expectation value of a Hermitian operator $A$ and its derivative with respect to the conjugate,
\begin{equation}\label{eq:exp_deriv}
    F[A] \equiv \trexpval{\m}{A\m},
    \qquad
    \frac{\partial F[A]}{\partial \m^*} = A\m .
\end{equation}
Every entry below follows from Eq.~\eqref{eq:exp_deriv} together with the chain rule: a terminal cost $\Phi = h(F)$ gives $\partial\Phi/\partial\m^* = h'(F)\,A\m(T)$, while a cost assembled from a time integral, $C = g(I)$ with $I=\int_0^T \phi\,dt$, contributes the source $g'(I)\,\partial\phi/\partial\m^*$, where $g'(I)$ is a scalar fixed by the completed forward pass. 

Table~\ref{tab:losses} collects the most common contributions. In the following, we work through three representative cases of increasing complexity.

\begin{table*}
\centering
\begin{ruledtabular}
\begin{tabular}{lccc}
Contribution & Loss $C$ & Boundary $\partial\Phi/\partial\m^*$ & Source $\partial\phi/\partial\m^*$ \\
\hline \\[-8pt]
State fidelity & $1 - \mathrm{tr}\left[{\m^\dagger (T)}{P^{\rm t}\m (T)}\right]$ & $-\,P^{\rm t}\m (T)$ & --- \\[6pt]
Average gate fidelity & $1 - \dfrac{1}{K}\displaystyle\sum_{k} \mathrm{tr}\left[{\m_{k}(T)}{P_k^{\rm t}\m_{k}(T)}\right]$ & $-\dfrac{1}{K}\,P_k^{\rm t}\m_{k}(T)$ & --- \\[10pt]
Running observable & $\displaystyle\int_0^T \trexpval{\m}{A\m}\,dt$ & --- & $A\m$ \\[10pt]
Nonlinear (terminal) & $h\!\big(F[A]\big)$ & $h'(F)\,A\m (T)$ & --- \\[6pt]
Nonlinear (running) & $G\!\big(\textstyle\int_0^T\phi\,dt\big)$ & --- & $G'(I)\,\partial\phi/\partial\m^*$ \\[6pt]
Control penalty$^{\,\dagger}$ & $\displaystyle\int_0^T \sum_k c_k\,u_k^2(t)\,dt$ & --- & --- \\
\end{tabular}
\end{ruledtabular}
\caption{Common loss-function contributions and the two derivatives needed to assemble them in LROC: the boundary condition for the adjoint state and the running source term. For the average gate fidelity, the boundary is applied per evolution $k$ and the gradients are summed. $^{\dagger}$Control penalties do not depend on $\m$ and bypass the adjoint entirely, contributing directly to the gradient $\partial C/\partial u_k^{(j)} = 2c_k\Delta t\,u_k^{(j)}$. In practice, however, they are enforced through the pulse parametrization and constraints (see Appendix~\ref{app:constraints}).}
\label{tab:losses}
\end{table*}

\subsection{State fidelity}

The simplest objective is the terminal state infidelity with objective $\ket{\psi^{\rm t}}$,
\begin{equation}
    C = 1 - \mathrm{tr}\left[{\m^\dagger(T)}{\ket{\psi^{\rm t}}\!\bra{\psi^{\rm t}}\m(T)}\right].
\end{equation}
Applying Eq.~\eqref{eq:exp_deriv} with $A=\ket{\psi^{\rm t}}\!\bra{\psi^{\rm t}}$ and $h(F)=1-F$ gives the adjoint boundary condition directly,
\begin{equation}
    \bm\lambda(T) = \frac{\partial \Phi}{\partial\m^*}
    = -\,\ket{\psi^{\rm t}}\!\bra{\psi^{\rm t}}\m(T),
\end{equation}
with no running source. The average gate fidelity is the immediate generalization: each of the $K$ 2-design states is propagated independently, their corresponding adjoint is initialized with $-K^{-1}\ket{\psi^{\rm t}}\!\bra{\psi^{\rm t}}\m_{k}(T)$, and the $K$ gradients are summed.

\subsection{Running cost: transmon readout}

Transmon readout combines several features at once: a running cost, two coupled trajectories, an outer nonlinearity, and a penalty. Each of these pieces reduces to a clean source. Two density matrices $\m_g$ and $\m_e$ are propagated from the qubit states $\ket{g}$ and $\ket{e}$, with cavity amplitudes $\alpha_q(t) = \trexpval{\m_q}{a\m_q}$ and pointer separation $\Delta\alpha = \alpha_g - \alpha_e$. Maximizing the signal-to-noise ratio corresponds to minimizing
\begin{equation}
    C_{\rm SNR} = \left(2\eta\kappa\, I\right)^{-1/2},
    \qquad I = \int_0^T |\Delta\alpha(t)|^2\,dt .
\end{equation}
The outer chain rule gives $g'(I) = -C_{\rm SNR}/2I$, a scalar fixed once the forward pass is complete. For the inner derivative, $\alpha_q$ is \emph{complex} because $a$ is non-Hermitian, so both $a$ and $a^\dagger$ appear,
\begin{equation}
    \frac{\partial |\Delta\alpha|^2}{\partial\m_g^*}
    = \big(\Delta\alpha^*\, a + \Delta\alpha\, a^\dagger\big)\m_g ,
\end{equation}
a manifestly Hermitian combination, with the opposite sign for the $e$ trajectory. The two adjoint equations are therefore sourced by
\begin{align}
    \frac{\partial\phi}{\partial\m_g^*}
    &= -\frac{C_{\rm SNR}}{2I}\big(\Delta\alpha^*\, a + \Delta\alpha\, a^\dagger\big)\m_g, \\
    \frac{\partial\phi}{\partial\m_e^*}
    &= +\frac{C_{\rm SNR}}{2I}\big(\Delta\alpha^*\, a + \Delta\alpha\, a^\dagger\big)\m_e,
\end{align}
and the total gradient sums the $g$ and $e$ contributions. A photon-number penalty $C_{n_{\rm max}} = \sum_{q=g,e}\mathrm{ReLU}( n_{\mathrm r, q} - n_{\max})$, with $n_{\mathrm r, q} = \trexpval{\m_q}{a^\dagger a\,\m_q}$, simply adds the source $\Theta(n_{\mathrm r, q} - n_{\max})\,a^\dagger a\,\m_q$ to each trajectory, with $\Theta$ the Heaviside function. Both terminal derivatives vanish: this is a pure running cost.

\subsection{Nonlinear objective: parity-check}

The parity-check objective of Eq.~\eqref{eq:qec_loss} is nonlinear in the state fidelities, demonstrating that the recipe is not restricted to objectives linear or quadratic in $\m$. For each state in the 2-design indexed by $k$, 
\begin{equation}
    C_{\rm QEC} = 1 - \frac{1}{K}\sum_{k=1}^K
    \Big(\sqrt{p_k^+} + \sqrt{p_k^-}\,\Big)^2,
\end{equation}
with branch fidelities $p_k^\pm = \trexpval{\m_{k}(T)}{P_k^{\pm}\m_{k}(T)}$ and projectors $P_k^{\pm} = \ket{\Psi_k^{\pm}}\!\bra{\Psi_k^{\pm}}$ onto the even/odd branches~\eqref{eq:qec_even_odd_branches}. Differentiating through the square root and the square yields a closed-form boundary condition in which the projectors acquire state-dependent weights,
\begin{align}
\begin{split}
    \bm\lambda_k(T) =& -\frac{1}{K}\Bigg[
    \Bigg(1+\sqrt{\tfrac{p_k^{-}}{p_k^+}}\Bigg) P_k^{+}
    \\
    &+ \Bigg(1+\sqrt{\tfrac{p_k^{+}}{p_k^{-}}}\Bigg) P_k^{-}
    \Bigg]\m_{k}(T).
\end{split}
\end{align}
The objective remains a terminal cost, so there is no running source. We note that the weights diverge when a branch fidelity vanishes [which can occur, since the branches~\eqref{eq:qec_even_odd_branches} are unnormalized and may be zero] and in practice we regularize $\sqrt{p}\to\sqrt{p+\varepsilon}$.

\medskip

Taken together, these applications show that any \emph{differentiable} functional
of the trajectory reduces to a boundary term, a running source, or both,
each obtained by a short application of Eq.~\eqref{eq:exp_deriv} and the
chain rule, while the dynamics adjoint of Appendix~\ref{app:formalism} is
reused unchanged. 

\section{Numerical details}\label{app:numerical}

\subsection{Practical implementation}\label{app:implementation}

The forward~\eqref{eq:nosse} and adjoint~\eqref{eq:adjoint_eom} equations are integrated with an adaptive fifth-order Tsitouras scheme (\texttt{Tsit5})~\cite{tsitouras_runge-kutta_2011, rackauckas_differential_2017} at fixed tolerances, decoupling the step count from $\NT$. The Gram matrix is Tikhonov-regularized, $\m^\dagger\m \to \m^\dagger\m + \delta\,\openone$ with $\delta = 10^{-8}\,\mathrm{tr}(\m^\dagger\m)/M$, before inversion. The initial state $\m_1$ places the initial physical state $\ket{\psi_0}$ in the first column with weight $\sqrt{1 - (M-1)\epsilon^2}$ and orthonormalizes $M-1$ auxiliary columns of weight $\epsilon \sim 10^{-5}$. Convergence is verified in both $M$ and $\epsilon$.

We implement the gradient in two interchangeable ways. In the first, the forward states $\{\mj\}$ are stored only at the $\NT + 1$ slice edges, at memory cost $\mathcal O(\NT \N M)$. Together with the trapezoidal error of Eq.~\eqref{eq:grape_gradient}, this sets the memory and error scaling of Figure~\ref{fig:memory_scaling}(c).

In the second, augmented formulation, the forward solution is stored densely and the adjoint coefficients, running-cost sources, and gradient integrands are evaluated at the interpolated state $\m(t)$. The gradient integrals~\eqref{eq:gradient_integral} are appended to the adjoint state and integrated by the same solver (for more details see Ref.~\cite{chen_neural_2019}). Gradient accuracy is then limited only by the solver tolerance at the cost of dense forward storage.

Multi-state objectives are threaded over the independent evolutions and the gradients summed. Parameters are then updated by gradient descent
or Adam~\cite{kingma_adam_2015} subject to the pulse constraints discussed in the following section. The stored-trajectory memory could be reduced by regenerating $\{\mj\}$ backward alongside the adjoint~\cite{gautier_optimal_2025}, but the contracting dissipative flow makes reverse integration unstable, requiring checkpoints spaced within $\sim 1/\Gamma_{\max}$; we do not use this strategy here, as the stored cost is already modest at low rank.

\subsection{Pulse constraints}\label{app:constraints}

Experimental pulses must vanish at the endpoints, respect bandwidth limits, and remain within amplitude bounds. Rather than penalizing violations in the loss, we enforce these constraints through the parametrization: the bare optimization variables $\ukj$ are mapped to the physical pulse by
\begin{equation}
    \mathbf u_{\rm phys} = W\, S\, \mathbf u,
\end{equation}
where $W = \mathrm{diag}(w_1,\dots,w_{\NT})$ is a fixed diagonal window enforcing smooth turn-on and turn-off, e.g. $w_j = \sin^2\!\big[\pi(j-\tfrac12)/\NT\big]$ or the Hann window $w_j = \tfrac12\big[1 - \cos\!\big(2\pi(j-1)/(\NT-1)\big)\big]$, and $S$ is a Gaussian smoothing operator limiting the pulse bandwidth, with entries
\begin{equation}
    S_{jj'} = \mathcal N_j^{-1}\,
    \exp\!\big[-(j-j')^2/2s^2\big], \qquad |j-j'| \le 3s,
\end{equation}
row-normalized so that $\sum_{j'} S_{jj'} = 1$, with kernel width $s$ in units of the control interval $\Delta t$ (boundary-clamped at the pulse edges). The gradient with respect to the bare variables follows from the chain rule, $\partial C/\partial \mathbf u = S^\top W^\top\, \partial C/\partial \mathbf u_{\rm phys}$, applied after the adjoint pass. Amplitude bounds are enforced by clamping after each optimizer step. Because the map is linear and fixed, these constraints add negligible cost and, unlike penalty terms, are satisfied exactly at every iteration.

\subsection{Comparison to automatic differentiation}\label{app:comp_ad}

The trade-off between hard-coded gradients and automatic differentiation (AD) in gradient-based quantum control is well characterized~\cite{leung_speedup_2017,abdelhafez_gradient-based_2019,goerz_quantum_2022,lu_optimal_2024}. In this section, we briefly summarize the key distinctions. 

The signature advantage of reverse-mode AD is that arbitrary, possibly non-analytic loss functionals can be differentiated without manual derivation~\cite{leung_speedup_2017, abdelhafez_gradient-based_2019}. Its memory cost was discussed in Sec.~\ref{sec:scaling}. Checkpointing strategies that trade this memory for repeated evolutions have so far been applied only to the full Lindblad equation, for which the scaling remains prohibitive~\cite{gautier_optimal_2025, gautier_high-fidelity_2024}. Systematic benchmarks find comparable runtimes across hard-coded, semi-automatic, and fully automatic gradients~\cite{lu_optimal_2024}, leaving memory as the decisive factor at scale.

A further distinction lies in the order of discretization and
differentiation. LROC is an \emph{optimize-then-discretize} scheme: the adjoint equation and gradient are derived at the continuous level [Eqs.~\eqref{eq:adjoint_eom_app}--\eqref{eq:dfdm}] and only then integrated numerically, so the computed gradient is a discretization of the exact continuous gradient. Reverse-mode AD is \emph{discretize-then-optimize}: the dynamics are first replaced by the finite composition of integrator steps, and this discrete map is differentiated exactly, yielding the exact gradient of the discretely evaluated loss. The two orders do not commute at finite $\Delta t$: their difference is the consistency error, which for piecewise-constant controls on the integration grid is the $\mathcal O(\Delta t^2)$ error of Eq.~\eqref{eq:grape_gradient} in the slice-edge implementation, and is reduced to the solver tolerance in the augmented one.


A final distinction is transparency and ease of use. Our method provides an explicit adjoint equation whose terms admit a geometric interpretation which can be extended by hand, as done here for trace-preserving dynamics and as required, e.g., for the tensor-network extension discussed in Sec.~\ref{sec:discussion}. On the other hand, AD can be used to evaluate any loss function without manually deriving the terminal adjoint and source terms. As complex-AD tooling for low-rank dynamics matures, we expect hybrid semi-automatic strategies~\cite{goerz_quantum_2022, lu_optimal_2024}, in which the analytic adjoint carries the low-rank propagation while AD supplies the cotangent of a non-analytic objective, to become the natural default.

\end{document}